\documentclass[journal = ancac3, manuscript = article]{achemso}

\usepackage{amsmath}
\usepackage{amssymb}
\usepackage{graphicx}
\usepackage{mathtools}
\usepackage{braket}
\usepackage{hyperref}
\usepackage{bbold}
\usepackage{calrsfs}
\usepackage{dsfont}

\usepackage{xcolor}
\usepackage{subcaption}

\newcommand{\RomanNumeralCaps}[1]
    {\MakeUppercase{\romannumeral #1}}
\DeclareMathOperator{\Tr}{Tr}

\title{Polaritonic Chemistry: Collective Strong Coupling Implies Strong Local Modification of Chemical Properties }

  \author{Dominik Sidler}
  \email{dsidler@mpsd.mpg.de}
  \affiliation{Max Planck Institute for the Structure and Dynamics of Matter and Center for Free-Electron Laser Science, Luruper Chaussee 149, 22761 Hamburg, Germany}
\alsoaffiliation{The Hamburg Center for Ultrafast Imaging, Luruper Chaussee 149, 22761 Hamburg, Germany}

 \author{Christian Sch\"afer}
  \email{christian.schaefer@mpsd.mpg.de}
  \affiliation{Max Planck Institute for the Structure and Dynamics of Matter and Center for Free-Electron Laser Science, Luruper Chaussee 149, 22761 Hamburg, Germany}
    \alsoaffiliation{The Hamburg Center for Ultrafast Imaging, Luruper Chaussee 149, 22761 Hamburg, Germany}
  \author{Michael Ruggenthaler}
  \email{michael.ruggenthaler@mpsd.mpg.de}
  \affiliation{Max Planck Institute for the Structure and Dynamics of Matter and Center for Free-Electron Laser Science, Luruper Chaussee 149, 22761 Hamburg, Germany}
  \alsoaffiliation{The Hamburg Center for Ultrafast Imaging, Luruper Chaussee 149, 22761 Hamburg, Germany}
 
  \author{Angel Rubio}
  \email{angel.rubio@mpsd.mpg.de}
  \affiliation{Max Planck Institute for the Structure and Dynamics of Matter and Center for Free-Electron Laser Science, Luruper Chaussee 149, 22761 Hamburg, Germany}
    \alsoaffiliation{The Hamburg Center for Ultrafast Imaging, Luruper Chaussee 149, 22761 Hamburg, Germany}
  \alsoaffiliation{Center for Computational Quantum Physics, Flatiron Institute, 162 5th Avenue, New York, NY 10010, USA}

\begin{document}

\begin{abstract}
Polaritonic chemistry has become a rapidly developing field within the last few years. A multitude of experimental observations suggest that chemical properties can be fundamentally altered and novel physical states appear when matter is strongly coupled to resonant cavity modes, i.e. when hybrid light-matter states emerge.
Up until now, theoretical approaches to explain and predict these observations were either limited to phenomenological quantum optical models, suited to describe collective polaritonic effects, or alternatively to \textit{ab initio} approaches for small system sizes.  The later methods were particularly controversial since collective effects could not be explicitly included due to the intrinsically low particle numbers, which are computationally accessible.
Here, we demonstrate for a nitrogen dimer chain of variable size that any impurity present in a collectively coupled chemical ensemble (e.g. temperature fluctuations or reaction process) induces local modifications in the polaritonic system. From this we deduce that a novel dark state is formed, whose local chemical properties  are modified considerably at the impurity due to the collectively coupled environment. Our simulations unify theoretical predictions from quantum optical models (e.g. formation of collective dark states and different polaritonic branches) with the single molecule quantum chemical perspective, which relies on the (quantized) redistribution of local charges. Moreover, our findings suggest that the recently developed QEDFT method is suitable to access these locally scaling polaritonic effects and it is a useful tool to better understand recent experimental results and to even design novel experimental approaches.
All of which paves the way for many novel discoveries and  applications in polaritonic chemistry. 
\end{abstract}


\maketitle
 


A multitude of fundamental experimental and theoretical breakthroughs have transformed polaritonic chemistry into a rapidly developing field over the past years. \cite{ebbesen2016,flick2017atoms,feist2017,ruggenthaler2018quantum}  The selective coupling of matter with photons in a cavity offers a broad range of novel experimental applications to modify and control matter properties on the nanoscale in an unprecedented way:
%
For example, single molecular optical and mechanical properties could by tuned by strong coupling in an optical resonator at cryogenic \cite{wang2017coherent,wang2019turning} or even room temperature.\cite{chikkaraddy2016single,ojambati2019quantum} 
Moreover, collective strong coupling was employed to control reaction rates experimentally\cite{hutchison2012,Thomas2016}, modify the critical temperature of superconductors\cite{thomas2019exploring} and to actually reach  Bose–Einstein condensation at room temperature.\cite{plumhof2014room}

Alongside with these impressive experimental progress, a plethora of theoretical models were developed to explain the experimental results. Particularly successful were collective models that either rely on on a (classical) mean-field description (e.g. Lorentz, Maxwell-Bloch)\cite{taflove2005computational,lugiato2015nonlinear} or collective quantum optical models, which also account for the quantization of the electromagnetic-field (e.g.  Tavis-Cummings or Dicke-model).\cite{tavis1968exact,dicke1954coherence} The latter models usually consist of a very large number $N$ of identical matter subsystems that couple to the quantized cavity photon field. Typically the matter subsystem  is only weakly interacting and they are approximated by a few energy levels. The fundamental interpretation of these quantum optics  models suggests that the presence of the cavity leads to the formation of bright, robust, collective quantum states i.e. polaritons, which are accompanied with a reservoir of collective dark states. \cite{tavis1968exact,gonzalez2016uncoupled,galego2017many}
Combining the models with characteristic experimental data\cite{chikkaraddy2016single,kleemann2017strong,ojambati2019quantum}  suggests that these polaritons extend to a mesoscopic scale and they can persist even at room temperature.
Typically, disorder of the molecular system is represented by a modification of the collective density of states (DOS) in the polaritonic ensemble,\cite{ribeiro2018polariton} whereas local molecular observables,   effectively remain unaltered by the dressing of a cavity.\cite{tavis1968exact,galego2017many} From this perspective, a perturbation of the molecular system can either break the coherence of the polaritons or the degeneracy of the dark states.\cite{agranovich2003cavity,litinskaya2006loss,agranovich2007nature,litinskaya2008propagation,ribeiro2018polariton} Additionally, originally bright, local, imperfections eventually turn dark by means of photonic intensity transfer, which reduces for example the broadening of the polaritonic spectra.\cite{manceau2017immunity,ribeiro2018polariton} Therefore,  in this picture induced polaritonic effects are explained by means of DOS reservoirs, which are then used to rationalize a multitude of experimental results in polaritonic chemistry (e.g relaxation dynamics, singlet-fission, energy-transfer).\cite{michetti2008simulation,martinez2018polariton,ribeiro2018polariton}

\begin{figure}[H]
\begin{subfigure}{1\textwidth}
\centering
    \includegraphics[width=1\linewidth]{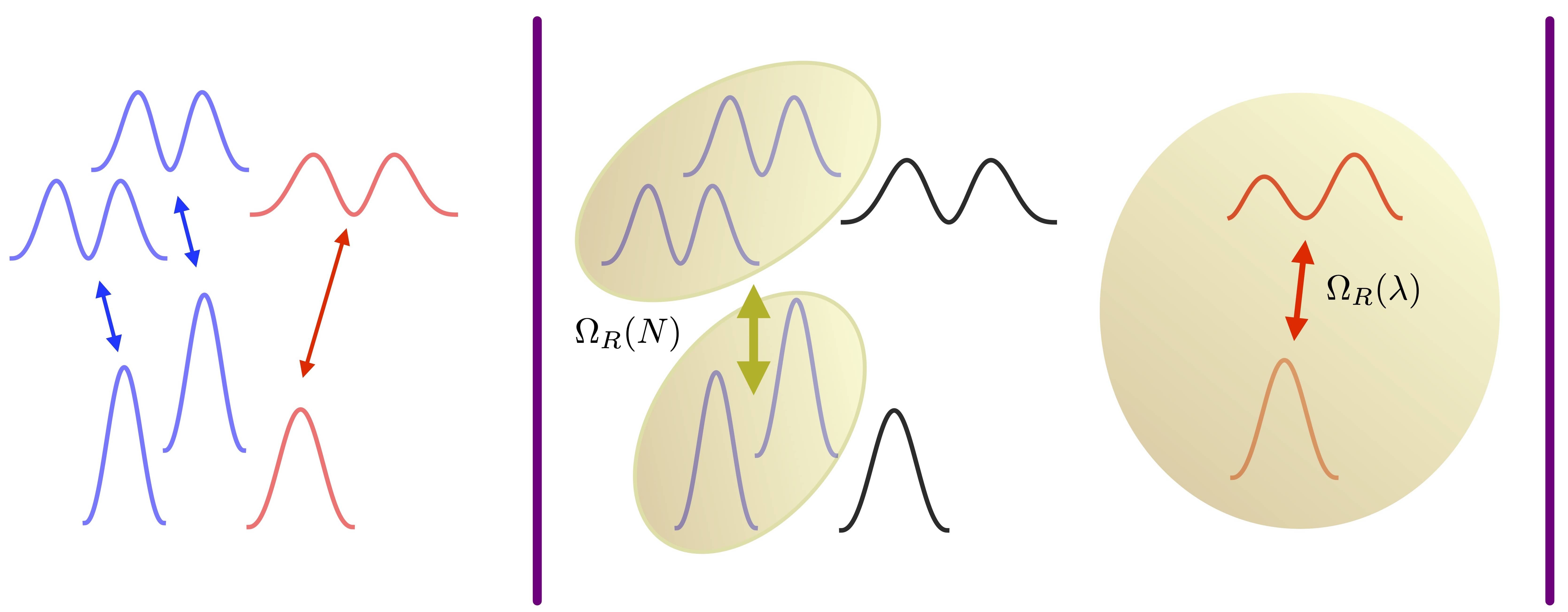}
\end{subfigure}
\caption{Sketch of different extreme theoretical views  on the polaritonic dressing of a homogeneous ensemble of $N$ molecules (blue) including one perturbation (red): Left: Uncoupled ensemble with different individual molecular excitations (indicated by arrows between the different ground an excited states). Middle: The extreme quantum optic perspective, which assumes the emergence of one collective macromolecule (olive) induced by the interaction with the cavity photons (cavity indicated by purple lines), whose Rabi splitting $\Omega_R$ scales with $\sqrt{N}$. Any perturbations are suppressed and considered dark (black). Right: The opposing single-molecule \textit{ab initio} quantum chemical perspective tries to explain polaritonic effects by strong local modifications of single molecular properties only. Typically, they are induced solely by a coupling parameter $\lambda$, which is determined by the collective environment.}
    \label{fig:extrem_cases}
\end{figure}

However, despite the success of these phenomenological models, experimental observations suggest that they may not yet capture all relevant aspects of polaritonic chemistry.\cite{thomas2016ground,hirai2020recent} In particular, the quantum optics models contradict traditional chemical intuition that chemistry is governed by local modifications of matter properties (e.g. charge and excitation transfers, conformational changes etc.). In addition, considering the tremendous challenges of creating macroscopic quantum states at significant temperatures in other areas of physics,\cite{frowis2018macroscopic} an explanation of the observed collective effects in terms of quantized mesoscopic states remains controversial.
%
One possible explanation for these discrepancies between experiment and theory may be attributed to the (over)-simplified matter description of these collective quantum models, which is necessary to make macroscopic $N$ regimes computationally accessible, but prohibits cavity induced local modifications of matter. In order to address this issue standard quantum optics models were extended to incorporate feedback loops with standard quantum-chemical methods, originally designed for molecular systems in the absence of a cavity.\cite{doi:10.1021/acs.jctc.7b00388,triana2019revealing}
Only very recently, rigorous \textit{ab initio} methods emerged, ~\cite{ruggenthaler2014quantum,ruggenthaler2015,schafer2018ab,mordovina2019polaritonic,csehi2019ultrafast,buchholz2019reduced} which allow to 
treat from first principles molecular systems and the quantized electromagnetic field.\cite{sentef2018cavity,schafer2019modification}
Based on this theoretical progress, polaritonic systems with real molecular constituents became within reach of DFT \cite{flick2018ab,flick2019light,doi:10.1080/00018732.2019.1695875,flick2020ab}, coupled cluster \cite{haugland2020coupled}  or even exact level of theory. \cite{sidler2020chemistry} Because these approaches have been mainly applied to single molecules coupled strongly to the quantized photon field, a complementary perspective of polaritonic chemistry has emerged. In contrast to the collective viewpoint, the collective ensemble is interpreted as an effective medium locally enhancing the coupling to the light field. A visualisation of two extreme theoretical viewpoints in polaritonic chemistry is given in Fig. \ref{fig:extrem_cases}. 

In the following work, we use such a novel \textit{ab initio} method to resolve this fundamental discrepancy between locally induced polaritonic effects and modifications of the collective polaritonic ensemble.  
For this purpose, we have designed a chain of nitrogen dimers within a cavity, which is chosen to meet the collective model assumptions as close as possible, but accessible with quantum-electrodynamical density-functional theory (QEDFT). With our simulations, we can confirm different collective effects that are predicted by  models (e.g. emergence of bright polaritonic branches, (quasi)-degenerated dark states and photonic intensity transfer).
However, when introducing a small perturbation of the bondlength in one of the dimers (e.g. due to the onset of a chemical reaction), we can demonstrate from first principles that collective coupling can indeed induce strong local modifications of molecular systems in the proximity of the (darkened) perturbation. This observation allows to embed local chemical modifications within the collective coupling context of polaritonic chemistry and it confirms that recently developed \textit{ab initio} methods are suitable to predict these locally induced modifications. Moreover, our findings highlight that standard collective models do not yet capture all relevant aspects of polaritonic chemistry.

\section{RESULTS/DISCUSSION}

In order to address the fundamental question of polaritonic chemistry, i.e. whether or not collective coupling imposes major local changes in chemical systems, we align a chain of nitrogen dimers in a cavity.  To investigate collective effects, we allow a variable chain length $N$ and we choose a large spacing (1.32 nm) between the dimers. The latter selection allows to mimic the dilute gas limit of Dicke-type models as close as possible in our \textit{ab initio} setup, since it avoids any overlap of the electronic structures.  This setting makes standard quantum optics models directly applicable, which either model inter-molecular forces by only dipole-dipole interactions or discard them altogether by non-interacting molecules.  Out of the $N$ dimers, we specifically perturbed the nuclear distance of one central dimer, which introduces an impurity in our chain. This localized impurity gives access to potential local scaling effects arising from collective strong coupling of the cavity tuned on resonance with $N-1$ unperturbed dimers (see Supporting Information (SI) for homogeneous chains). Naturally, it is reasonable to assume impurities in a real chemical system. Typically, they may arise from thermal fluctuations in a homogeneous setup (in our case the bondlength of one dimer) or they are already predetermined by the inclusion of a solute in a homogeneous solvent environment.


For the fundamental quantum mechanical description of light-matter interaction in a cavity, our simulations are based on the Pauli-Fierz (PF) Hamiltonian in the long-wavelength limit. It assumes the following form for one cavity mode in dipole approximation with atomic units:\cite{ruggenthaler2014quantum,Schaefer2020relevance} 
\begin{equation}
\hat{H} = \sum_{k} \frac{\bold{\hat{p}}_k^2}{2} +\sum_{k<l}\frac{1}{|\bold{\hat{r}}_l-\bold{\hat{r}}_k|}+\sum_k V_{\mathrm{ext}}(\bold{\hat{r}}_k)+\frac{1}{2}\bigg[\hat{p}^2+\omega_c^2\bigg(\hat{q}-\frac{\boldsymbol{\lambda}}{\omega_c}\cdot \hat{\bold{R}}\bigg)^2\bigg],
\label{eq:hamiltonian}
\end{equation}
%
where the Born-Oppenheimer approximation was employed for the nuclear degrees of freedom.
The usual momentum and position operators of the electrons $k$ are $\bold{\hat{p}}_k,\bold{\hat{r}}_k$ .  The external potential of the nuclei is given by $V_{\mathrm{ext}}$ and $\hat{\bold{R}}=-\sum_k \hat{\bold{r}}_k$ corresponds to the electronic dipole operator. Quantization of light is described by the displacement coordinate $\hat{q}$ and its conjugate momentum operator $\hat{p}$, with associated mode frequency $\omega_c$ and coupling $\boldsymbol{\lambda}=\sqrt{\frac{4\pi}{ V}}\bold{e}_z$, which determines the transversal light-matter interaction for a cavity with an effective mode volume V.
To solve the stationary eigenvalue problem of the PF Hamiltonian numerically, we rely on its reformulation in the QEDFT framework.\cite{ruggenthaler2014quantum}
More specifically we use the light-matter linear response framework introduced in Ref. \citenum{flick2019light}. It gives access to spatially resolved ($\bold{r}$) electronic transition densities $\rho_{0j}(\bold{r})=\delta_{\bold{r},\bold{r}^\prime}\bra{j}\hat{\rho}(\bold{r},\bold{r}^\prime)\ket{0}$, for an excitation frequency $\omega_j$ with eigenfunction $\ket{j}$ of our cavity dressed system.\cite{li2011time}  The reduced one-particle density matrix operator is defined as $\hat{\rho}$.  Based on the transition densities, one has access to local observables such as transition dipole moments $z_{0j}^i= \Tr(z\rho_{0j})$ or oscillator strengths $f^{zi}_{0j}$  of dimer $i$ along $z$ due to Fermi's golden rule (see Ref. \citenum{flick2019light}). 

\begin{figure}
\begin{subfigure}{1\textwidth}
\centering
    \includegraphics[width=1\linewidth]{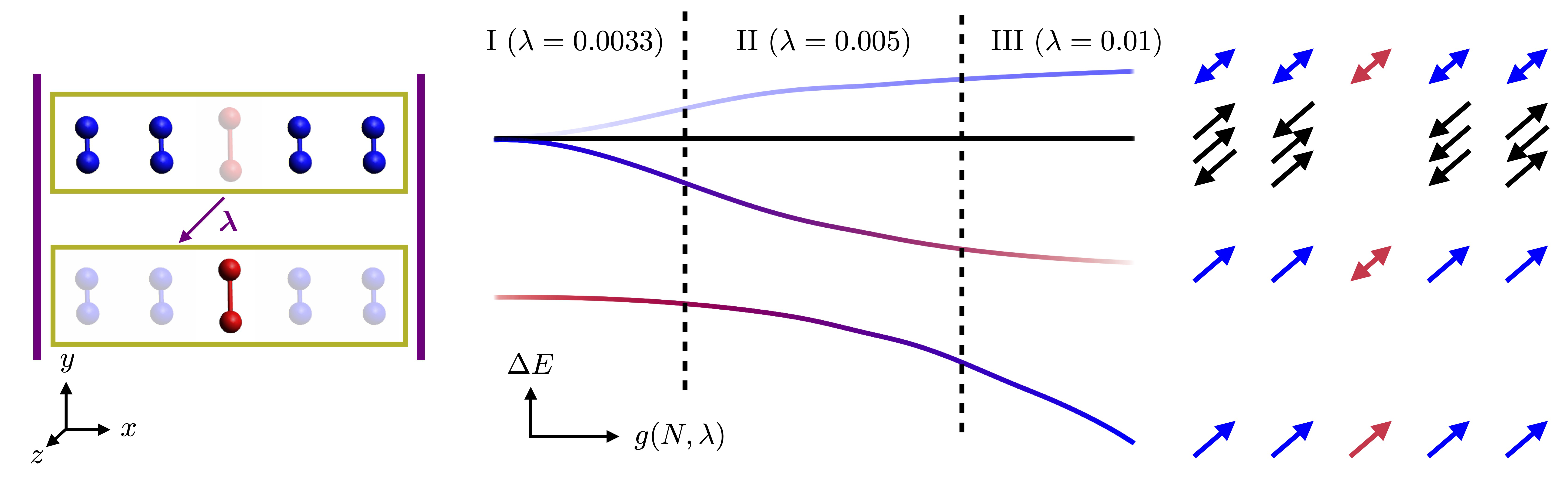}
\end{subfigure}
\caption{Left: Illustration of perturbed (red) nitrogen dimer chain of variable size with respect to the unperturbed dimers (blue). The polarization direction of the cavity mode in $z$-direction is indicated by $\boldsymbol{\lambda}$. Middle: Sketch of the energy splitting into collective upper, middle and lower polaritonic branches with respect to different coupling strengths $g$. Dark states are visualised in black. The color indicates the mainly contributing dimers  and the brightness relates to the oscillator strenghts of the underlying absorption spectra. Right: Potential transition dipole $z_{oj}$ alignment  patterns for branch $E_j$. Double arrows indicate a sign flip of the local transition dipole, which can occur at a certain coupling strength.}
\label{fig:res_overview}
\end{figure}

\begin{figure}
\begin{subfigure}{1\textwidth}
\centering
    \includegraphics[width=0.6\linewidth]{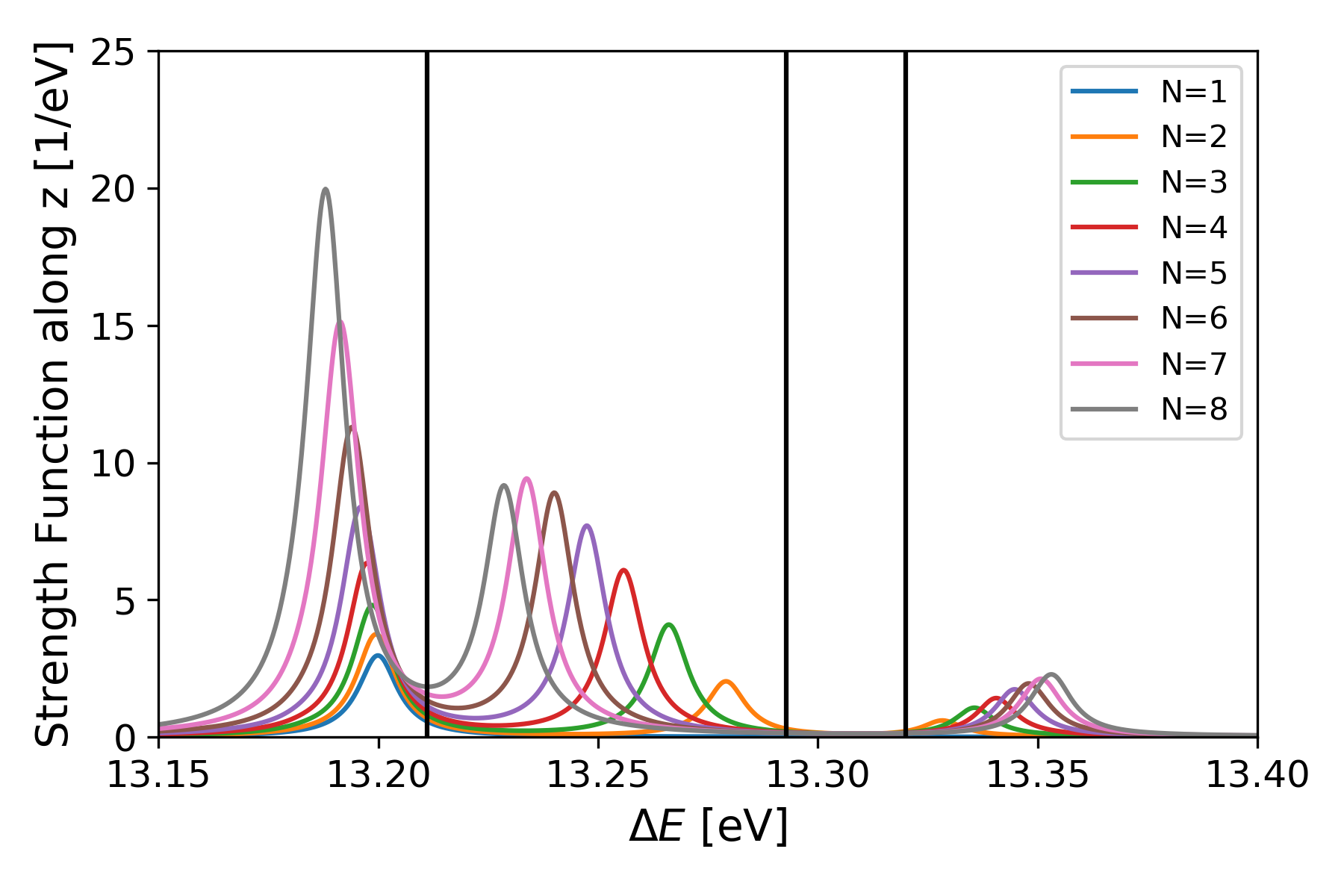}
\end{subfigure}
\caption{Lorentz-broadened collective spectra in coupling regime \RomanNumeralCaps{2} ($\lambda=0.005$), which is derived from the $z$-contribution of the oscillator strength $f^{zi}_{0j}$, i.e. with respect to the polarisation of the cavity. The formation of different polaritonic branches (separated by vertical black bars) can be observed with respect to different chain lengths $N$. From left to right: lower, middle, dark and upper polaritonic branch.  }
\label{fig:spec_broaden}
\end{figure}

Our simulations of the perturbed chain can reproduce the typical collective observables also captured by collective models. An overview of the main results is given in Fig. \ref{fig:res_overview}, which is supported by detailed spectral data in the Supporting Information. They reveal the formation of three different polaritonic branches (lower, middle, upper) in the vicinity of the cavity resonance energy, which are accompanied by (almost) degenerat dark states (see Fig. \ref{fig:spec_broaden}).  The appearance of a distinguishable middle and lower polaritonic state is caused by the specific choice of our perturbed dimer. It is designed to have its first uncoupled resonance  $\hbar\omega_p=13.206$ eV just below the unperturbed dimer resonance at $\hbar\omega_u=13.309$ eV. 
Depending on the magnitude of the Rabi split with respect to $\Delta\omega=\omega_u-\omega_p$, three different coupling regimes $\{$\RomanNumeralCaps{1},\RomanNumeralCaps{2},\RomanNumeralCaps{3}$\}$ can be assigned for the lower and middle polariton (see Fig. \ref{fig:res_overview}):
\renewcommand{\labelenumi}{\Roman{enumi}}

   \begin{enumerate}
     \item The Rabi splitting defined by the bright upper and middle polaritonic branch is too small to reach significant energetic overlap with our perturbation. Therefore, the perturbed excitation is energetically well separated from any collective effects. This implies that the chemistry at the (excited) perturbation is accessible with standard quantum chemical methods (e.g. DFT), whereas standard quantum optical models will be directly applicable to the dressed dimers.
     \item The Rabi splitting reaches considerable overlap with our perturbation. Avoided crossing between middle and lower polariton affects frequency and magnitude of their global spectral properties. Collective effects start to influence local chemical properties of the perturbation, and in principle a fully coupled \textit{ab initio} description of the entire ensemble would be required (e.g. QEDFT). This regime will be the main focus of the following article. 
     \item The Rabi energy split is considerably larger than $\Delta \omega$, which effectively darkens the middle polariton branch and faints the upper polaritonic branch of our setup. Therefore, in regime \RomanNumeralCaps{3} only one bright collective polaritonic state (lower) survives, with intermediate dark and relatively faint upper polaritonic branches. From a spectral point of view, one might expect that standard quantum optics models are suitable again to describe this regime with only two relevant polaritonic branches (lower, upper). %
 \end{enumerate}


\begin{figure}
\begin{subfigure}{1\textwidth}
\centering
    \includegraphics[trim=60 0 0 0, clip, width=1\linewidth]{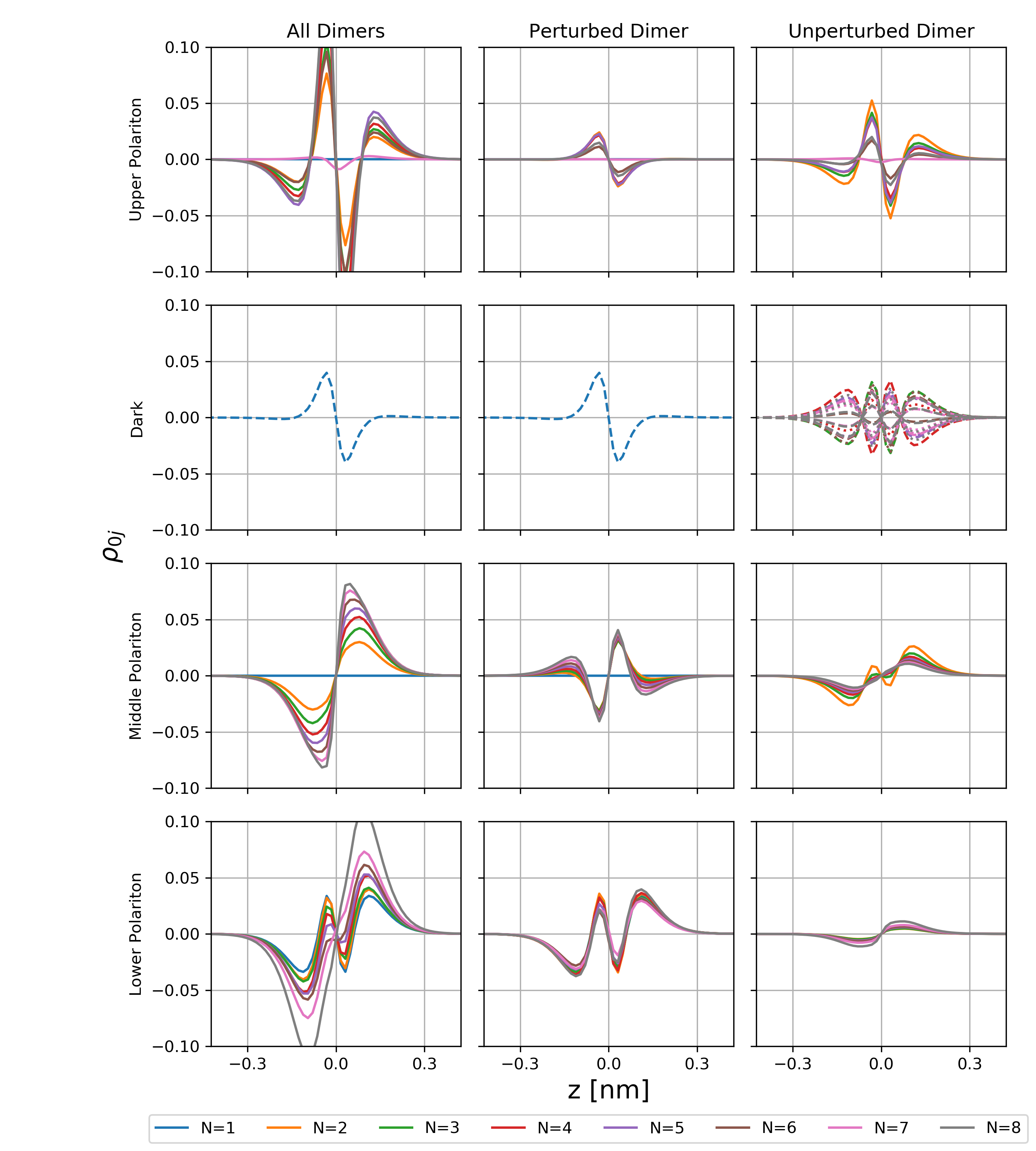}
\end{subfigure}
    \caption{Globally (left column) and locally resolved transition densities projected onto the $z$-axis for different chain lengths $N$ in coupling regime \RomanNumeralCaps{2}. For each of the four energy windows (rows), integrated quantities are displayed, except for the dark states. The integration cleans the data and contributes only very little to the overall results,except for $N=7$, i.e. in the vicinity of the observed splitting of the upper polaritonic branch. }
\label{fig:middle_p_scal}
\end{figure}

\begin{figure}
\begin{subfigure}{1\textwidth}
\centering
    \includegraphics[width=0.4\linewidth]{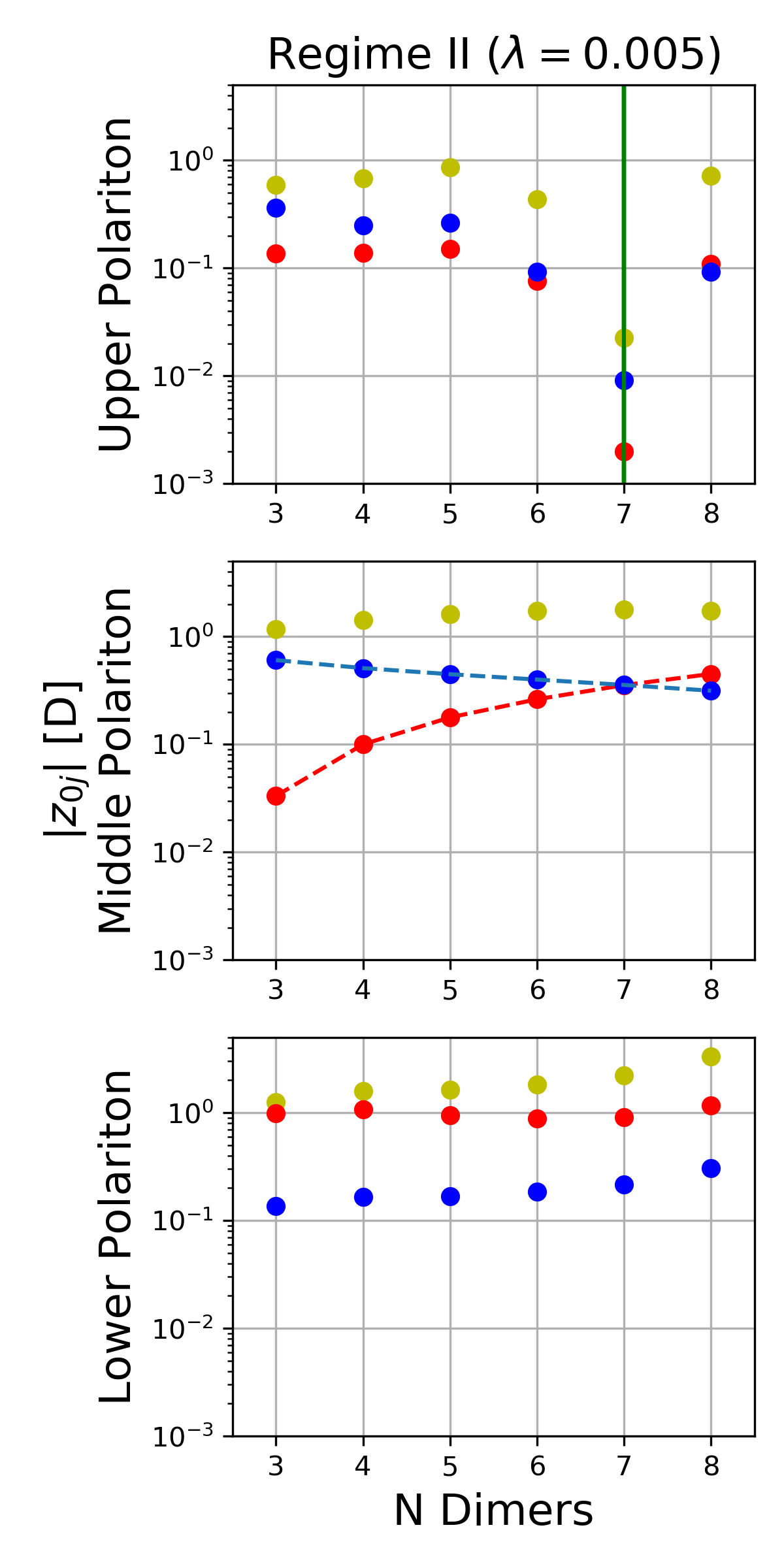}
\end{subfigure}
    \caption{Collective (yellow) and local transition dipole moment scaling with respect to different chain lengths $N$ for the perturbed (red) and unperturbed (blue) dimers in coupling regime \RomanNumeralCaps{2}. Special cases $N=\{1,2\}$ are excluded for the sake of clarity (either does not include an unperturbed dimer or no dark states can form). Opposing local scaling behaviour is indicated by blue and red lines in the middle polaritonic branch. The vertical green line indicates the observed splitting of the upper polaritonic branch, with locally canceling transition dipole moments when building a superposition of both states. See SI for other coupling regimes. }
\label{fig:dipole_scale}
\end{figure}

However, the aforementioned global spectral view on our system does not tell the whole story. The calculated transition densities give access to the local scaling behavior of the molecular charge distribution dressed in a cavity. For example, Fig. \ref{fig:middle_p_scal} demonstrates for the middle polariton that the local transition density of the perturbed dimer effectively scales up with increasing number of unperturbed dimers, while the converse is true for the $N-1$ unperturbed dimers. This enhancement of local charge re-distributions implies that here the chemical properties of the perturbed dimer are effectively modified on a local scale, whereas spectroscopically the middle polaritonic branch  darkens (regime \RomanNumeralCaps{3} in SI) as predicted by quantum optics models.\cite{manceau2017immunity} Indeed, the local (up)-scaling at the perturbation can be measured in terms of transition dipole moments (see Fig. \ref{fig:dipole_scale}) across all three coupling regimes in our simulations (Fig. S9 in SI). In regime \RomanNumeralCaps{1}, it may still be considered negligibly small, but once sufficient energetic overlap between middle and lower polariton is reached, local chemical properties can be affected substantially. Moreover, the collectively induced local modifications even persist in regime \RomanNumeralCaps{3}, where the middle polariton turns dark! From this we conclude that a novel dark state is induced by local perturbations, which has locally modified chemical properties due to the collective strong coupling. 

Rephrasing it differently, we discovered that local perturbations can impose locally opposite scaling behavior than predicted from  impurity free quantum optics models, which typically claim the absence of any local modifications. 
Therefore, our results imply that local chemical properties are modified by the fact that the perturbation is embedded in the collective state. This observation does not contradict collective predictions from phenomenological models though, indeed it unifies the  collective perspective predominant in the polaritonic physics community with the spatially resolved view of traditional chemistry as we will show subsequently.  

Notice that the scaling of the transition dipole moments in Fig. \ref{fig:dipole_scale} reveals an interesting feature in the upper polaritonic branch of our nitrogen chain, as there is a sudden decay at $N=7$ in coupling regime \RomanNumeralCaps{2}. The reason for this observation is  the specific transition density distribution of the nitrogen dimers, which have a locally decreasing transition dipole moment $z_{0j}$ with respect to the coupling strength. In principle,  $z_{0j}$  would vanish at a certain point. However, a more refined look at our results reveals that the upper polariton is indeed split into two energetically close subbranches, with opposite sign for the  underlying dipole transition elements (see Fig. S8 in SI). Entering into this splitting regime may trigger a multitude of  collective physical effects, which will be subject of future research.

\begin{figure}
\begin{subfigure}{1\textwidth}
\centering
    \includegraphics[trim=60 0 0 0, clip, width=1\linewidth]{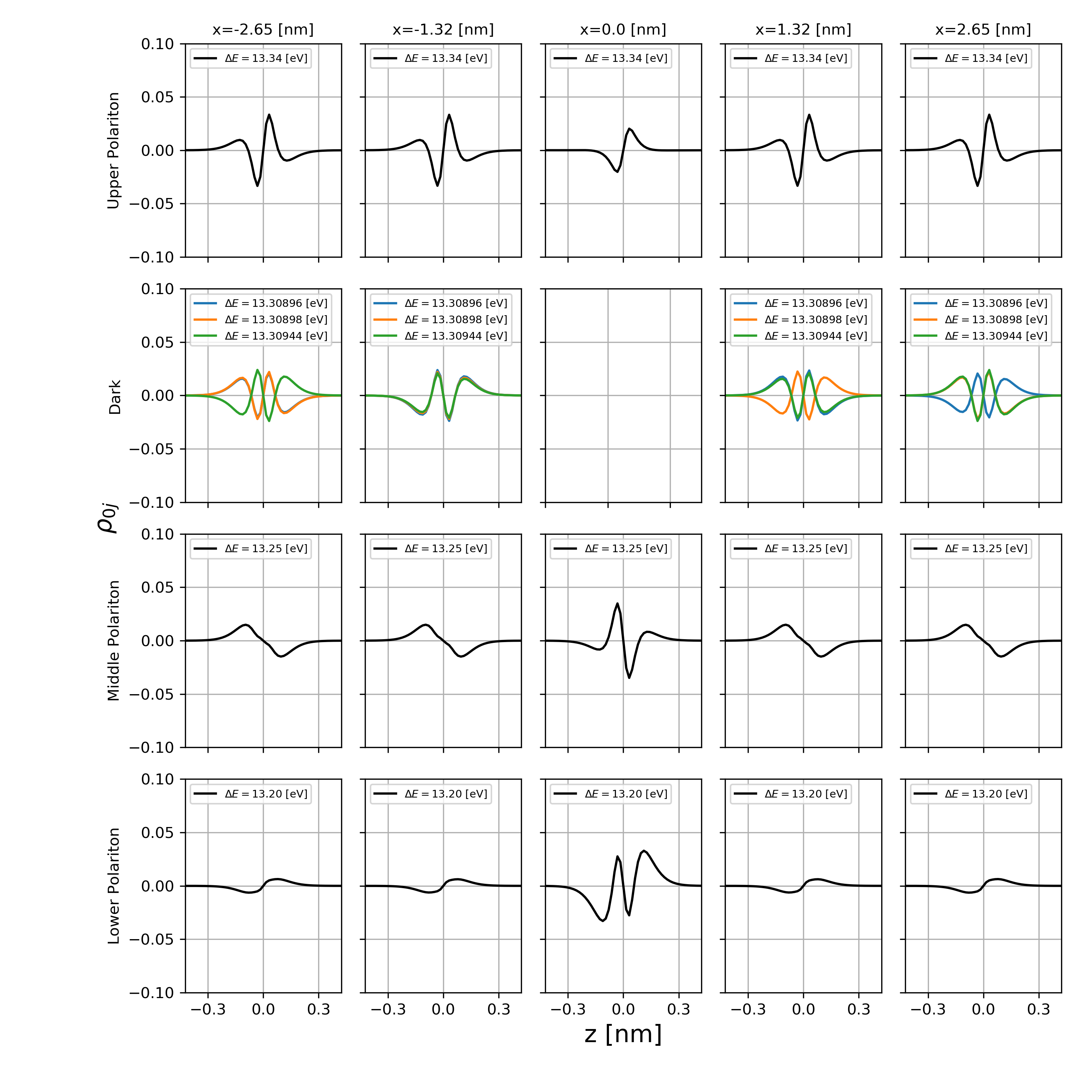}
\end{subfigure}
    \caption{Locally resolved transition densities of each dimer projected onto the $z$-axis for $N=5$ and coupling regime \RomanNumeralCaps{2}. For each of the three polaritonic branches (lower, middle, upper), the sum over all transition densities is taken from their respective energy windows. }
\label{fig:td_loc_Nd_5}
\end{figure}

An especially instructive case is given by $N=5$ for regime \RomanNumeralCaps{2} in Fig. \ref{fig:td_loc_Nd_5}. 
Here we see the locally projected transition densities of each individual dimer with respect to the $z$-axis for all three polaritonic branches as well as for the dark states. It nicely illustrates the relative orientation of the local transition dipole moments with respect to each other for the different polaritonic branches, resulting in their respective ordering in energy space (see also Fig. \ref{fig:res_overview}). Furthermore, this figure illustrates that boundary effects do not hamper our results for the chosen setup  and that for a given number of dimers, all unperturbed dimers behave identically. This property is crucial for the generalisation of our results with respect to collective coupling. Clearly, our \textit{ab initio} calculations are restricted to relatively few dimers only, while phenomenological models do not have this limitations. However, due to the absence of any finite size effects in our small ensemble, we deduce that we can in principle extrapolate to arbitrary chain lengths in the vicinity of a perturbation, i.e.  by a suitable choice of $\lambda$ and a relatively small number of explicitly simulated dimers (see Fig. S10 in SI). This makes the working assumption of current ab-initio simulations reasonable that collective strong coupling can be approximated by local strong coupling.  However, the accurate approximation of the local transition densities in all polaritonic branches at once remains an open question. If possible, it may require careful parameter tuning, which 
 becomes particularly relevant for defects if one leaves our dilute gas limit, and enters chemically more realistic regimes, where electron-overlap starts to play a significant role.


To our knowledge, our simulations also contain the first \textit{ab initio} description of collective dark states in polaritonic chemistry. As can be seen in Fig. \ref{fig:td_loc_Nd_5} for $N=5$, the four unperturbed local transition dipole moments anti-align in three different patterns along the chain ($x$-axis), with zero net transition dipole moment, which collectively makes them dark (see Fig. \ref{fig:res_overview}). However, at the same time, the three distinguishable dark state pattern effectively induce small local changes in the dipole transition densities, which lifts the degeneracy of the collective dark states in our fully interacting \textit{ab initio} perspective. This in turn means that collective dark states do not only affect the DOS of the collective ensemble, but also local chemical properties can indeed be modified by different  dark state alignment patterns. Notice that this property remains preserved for all simulations of collective dark states, which consist of an even number of unperturbed dimers, i.e. when  the local transition dipole moments can cancel each other exactly without finite size effects (see e.g. Fig S11 for $N=7$ in the Supporting Information).  
Consequently, collective dark state modified local chemical processes will be challenging to describe with \textit{ab initio} simulation methods due to their intrinsic system size limitations.
However, for the description of the discovered dark states with induced local modifications at defects, they are the method of choice. 

In addition, our simulations also indicate the existence of opposite local scaling behaviour not only for dark, but also for bright states if the cavity is de-tuned. As shown in the Supporting Information, setting the cavity in resonance with the perturbed dimer, instead of the unperturbed ones, does not fundamentally alter any qualitative prediction we made so far with one major exception: It appears that under these conditions, locally induced modifications of the electronic structure may start to play a significant role not only for the darkened middle polariton, but also for the bright upper polaritonic branch (see Fig. S17 in SI).  Certainly, when strongly detuned from any matter excitation, no collective Rabi splitting or local modifications are observed. 


\section{CONCLUSION}

In summary, we  demonstrated by \textit{ab initio} simulations that a collective strong coupling induces strong local modifications in polaritonic chemistry as soon as there is any impurity present in the ensemble. Overall our observation unifies predictions from collective models in quantum optics (e.g. formation of global dark states and different polaritonic branches) with the quantum chemical perspective, which relies on local redistribution of charges. This is in line with very recent, extensions of collective models\cite{schutz2020ensemble,hagenmuller2020adiabatic} which also suggest local strong coupling for defects that are induced by a weakly coupled collective molecular state. However, the predicted emergence of a Rabi splitting at the defect, linearly scaling with $N$, could not be confirmed for our chosen \textit{ab initio} setup of nitrogen dimers. It is characterized by the complex interplay of the three polaritonic branches instead, which leads to the emergence of a novel dark state when increasing the collective coupling strength. However, co-existing collective polaritonic properties remain qualitatively invariant. These novel dark states feature the locally induced modifications, which arise from the embedding of molecular impurities in a collective environment.  
Overall, our results suggest that the recently developed QEDFT method is suitable to access these local polaritonic effects and it is a useful tool to better understand recent experimental results and to even design novel experimental approaches.

Indeed, the structure of the local modifications induced by the collectively coupled ensemble suggest a more direct and simple interpretation of the observed effects in polaritonic chemistry. As demonstrated for the scaling of the transition dipoles, the cavity can polarize the ensemble within a certain frequency window such that the perturbed dimers feels a strong local field. This is not unlike local-field effects in solid state physics and it does not necessarily imply a quantum effect of the photon field.
In other words, it raises the question if also in polaritonic chemistry, quantum mechanical effects play only a significant role on a local scale, as suggested by recent experimental results with classical plasmonic arrays\cite{baranov2020ultrastrong}, while classical physics takes over on the larger scale.
To shine further light on this pivotal question, recently developed experimental techniques to measure electromagnetic fields on an atomistic scale\cite{peller2020} may offer a promising tool. 

Overall, our findings open the door to include a multitude of quantum chemical concepts and methods into the polaritonic chemistry context, which were successfully developed and applied over the past decades. For example, one can give up our initial dilute system assumptions and start to investigate more realistic and complex chemical setups by including for example temperature, electron overlap, different orientations, solvent effects, driving laser fields and much more.\cite{alonso2012combination,agostini2020tddft} 
This paves the way to many novel discoveries and applications based on \textit{ab initio} polaritonic chemistry methods. All of which nurtures the hope  that by collective strong coupling, unprecedented local control of chemical processes may become within reach.

Aside from this, the observed collective sign change of the local dipole transition moments opens novel ground for future investigations. The observed collective sign flip implies that the light matter coupling suddenly ceases to exist up to the first order expansion in $\lambda$. Hence the  system undergoes a collective  phase transition at this point, which potentially triggers a multitude of unique dynamic, radiative and higher order coupling effects that fundamentally alter polaritonic properties on an atomistic as well as on a mesoscopic scale. The nature and properties of this phase transition will be topic of future research.

\section{METHOD}
QEDFT and DFT simulations were performed with OCTOPUS \cite{tancogne2020octopus} using the light-matter linear response framework for cavities, which was introduced in Ref.\citenum{flick2019light}.
For the ground state calculation, the optimized effective potential method with Krieger-Li-Iafrate (KLI) approximation\cite{krieger1990derivation,krieger1992construction,krieger1992systematic} was employed. Perturbations of the groundstate density due to the photon field were not considered for consistency with usual quantum optics model assumptions.

Each individual nitrogen dimer was aligned along the y-axis, whereas the chain extends along the x-axis with a cavity coupled along the z-axis (other cavity orientations are discussed in the SI).
The dimers were separated $1.32$ nm  with respect to their nearest neighbors in order to mimic (quasi) independent subsystems, which are mainly coupled via the cavity mode (dilute gas limit). This clearly avoids any density overlap in the chosen energy range of interest without the presence of a cavity. Simulations were performed for $N\in\{1,..,8\}$ with $N-1$ unperturbed dimers and a single perturbed dimer, which is always located at the center of the chain. The geometry of the ground-state nitrogen dimers was minimized in OCTOPUS reaching a nuclei distance of $107.9$ pm, for the unperturbed dimers.
The nuclear distance of the perturbed dimer was set to $110.5$ pm, which shifts the lowest lying excitation in the  absorption spectrum from $\hbar\omega_u=13.309$ eV for the unperturbed dimer to $\hbar\omega_p=13.206$ eV for an uncoupled system (i.e. for $\lambda=0$).
Core electrons of the nitrogen atoms were approximated with the standard Troullier-Martins pseudopotential, included in the OCTOPUS distribution.
A  real space Cartesian mesh with $15.9$ pm grid spacing was used for the wave-function representation with minimum boxshape radius $1.06$ nm. For derivative calculations, the standard OCTOPUS fourth order star stencils scheme was applied.

During the groundstate calculation, additional 120 unoccupied extrastates were minimized in the SCF cycles, which are necessary to extract the excited states in the linear response framework. One cavity mode was considered in the calculations. It was tuned to the calculated resonance frequency $\omega_c=\omega_u$, which corresponds to the first excitation of the unperturbed nitrogen dimer in the absence of a cavity. To give access to the three different scaling regimes $\{$\RomanNumeralCaps{1},\RomanNumeralCaps{2},\RomanNumeralCaps{3}$\}$, three different cavity coupling parameters were considered, which were set by $\lambda\in\{\frac{2}{3},1,2\}\times 0.005$. Additional simulations were conducted in regime \RomanNumeralCaps{2} with a (de)-tuned cavity on resonance with the perturbed dimer, i.e. $\omega_c=\omega_p$ (see SI).
Local observables (e.g. dipole transition moments, transition densities) were extracted from the spatially resolved output of the diagonal elements of the transition density matrix\cite{li2011time}, using a customized python script. The oscillator strenghts were Lorentz broadened with a full width at half maximum of $5.4$ meV to derive the spectral strength function. For illustration purposes, locally resolved transition densities were projected onto the $z$-axis, i.e. in agreement with the chosen cavity orientation  $\boldsymbol{\lambda}\parallel \bold{e}_z$. Moreover, only significant transition densities are displayed with a magnitude larger than $\pm 0.005$ for the individually resolved dimers and for the dark states. In contrast, to clear the data, the transition densities with respect to $N$ were integrated over the energy windows, which can be attributed to the lower, middle and upper polaritonic branches respectively. Analyses showed little differences compared with solely considering the respective main contributors, except in the vicinity of the observed splitting of the upper polaritonic branch for $N=7$ with $\lambda=0.005$, which will be subject of future investigations.

\begin{acknowledgement}
The authors thank Johannes Flick, Enrico Ronca and Claudiu Genes for helpful discussions and critical comments. This work was made possible through the support of the RouTe Project (13N14839), financed by the Federal Ministry of Education and Research (Bundesministerium für Bildung und Forschung (BMBF)) and supported by the European Research Council (ERC-2015-AdG694097), the Cluster of Excellence "CUI: Advanced Imaging of Matter" of the Deutsche Forschungsgemeinschaft (DFG), EXC 2056, project ID 390715994 and the Grupos Consolidados (IT1249-19). The Flatiron Institute is a division of the Simons Foundation.

\end{acknowledgement}

\begin{suppinfo}
Additional simulation data is provided for coupling regime \RomanNumeralCaps{1} in Figs. S1 - S5, for \RomanNumeralCaps{2} in Figs. S6 - S17 and for \RomanNumeralCaps{3} in Figs. S18 - S21. Generally, the data contains spectral data for chains without perturbations as well as spectra for different cavity polarisation along the $x$ and $y$-axis. For regime \RomanNumeralCaps{1} and \RomanNumeralCaps{3} local transition densities and transition dipole moments are displayed similar to this article. Moreover, in Fig. S12 - S17 our data analysis is extended to coupling regime \RomanNumeralCaps{2}, but with the cavity de-tuned to the perturbed dimer, i.e. $\omega_c=\omega_p$.

\end{suppinfo}
\bibliography{manuscript}
\end{document}





\maketitle
 

\section{Simulation Results for Coupling Regime \RomanNumeralCaps{1} ($\lambda=0.0033$)}

\subsection{Absorption Spectra for Cavity in Resonance with Unperturbed Dimers}

\begin{figure}[H]
\begin{subfigure}{.8\textwidth}
\centering
    \includegraphics[width=0.52\linewidth]{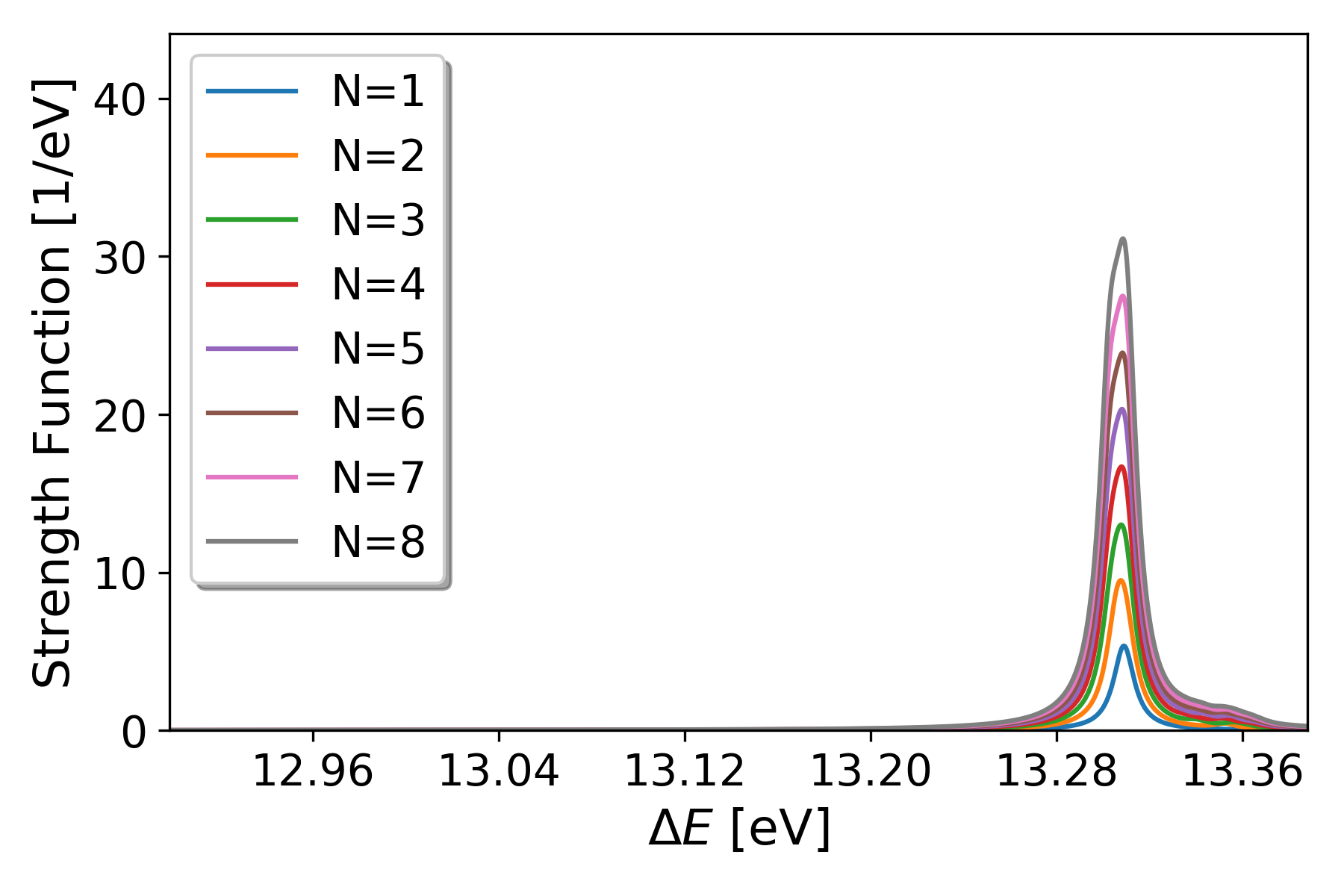}
    \label{fig:}
\end{subfigure}
\begin{subfigure}{.8\textwidth}
\centering
    \includegraphics[width=0.52\linewidth]{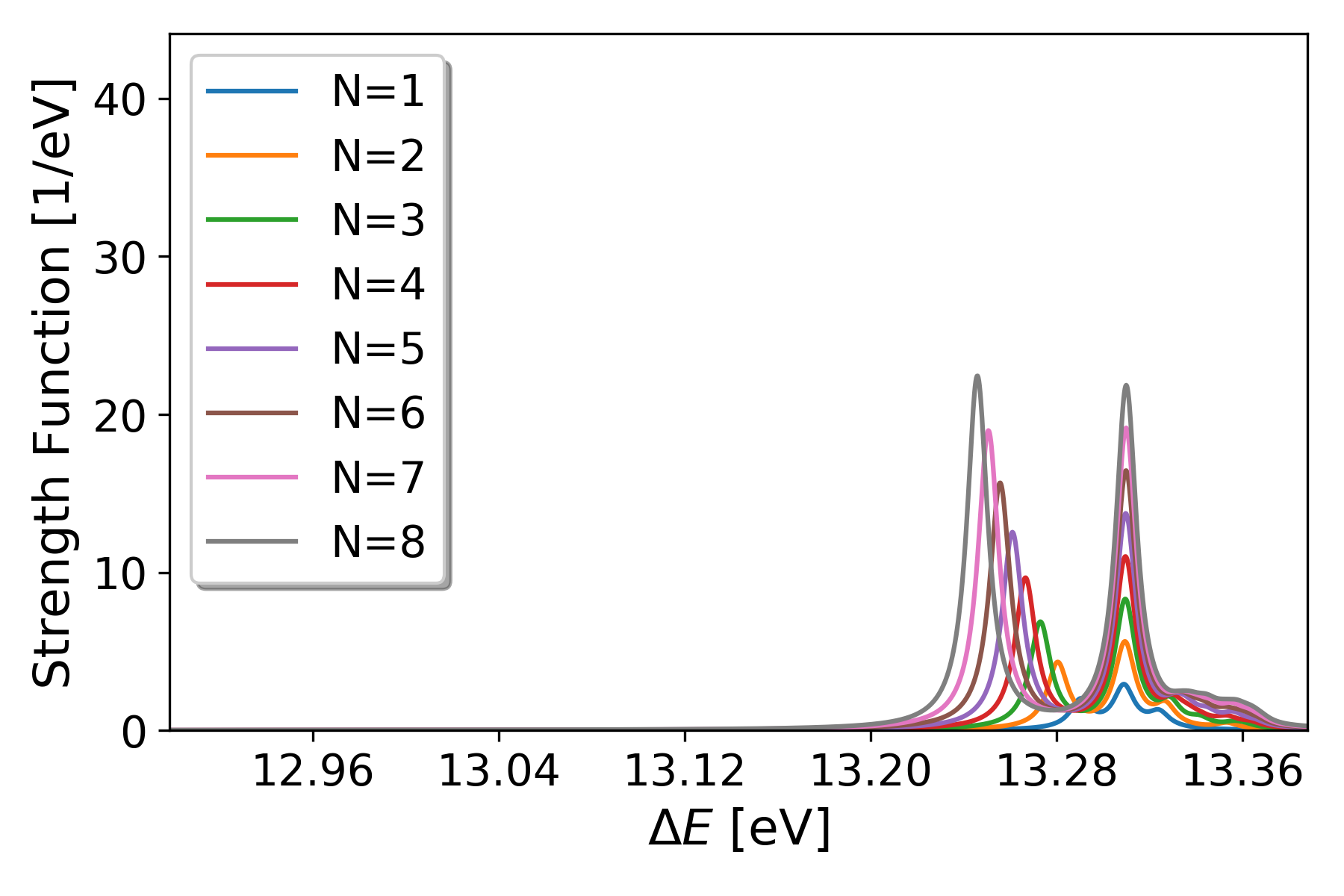}
    \label{fig:}
\end{subfigure}
\begin{subfigure}{.8\textwidth}
\centering
    \includegraphics[width=0.52\linewidth]{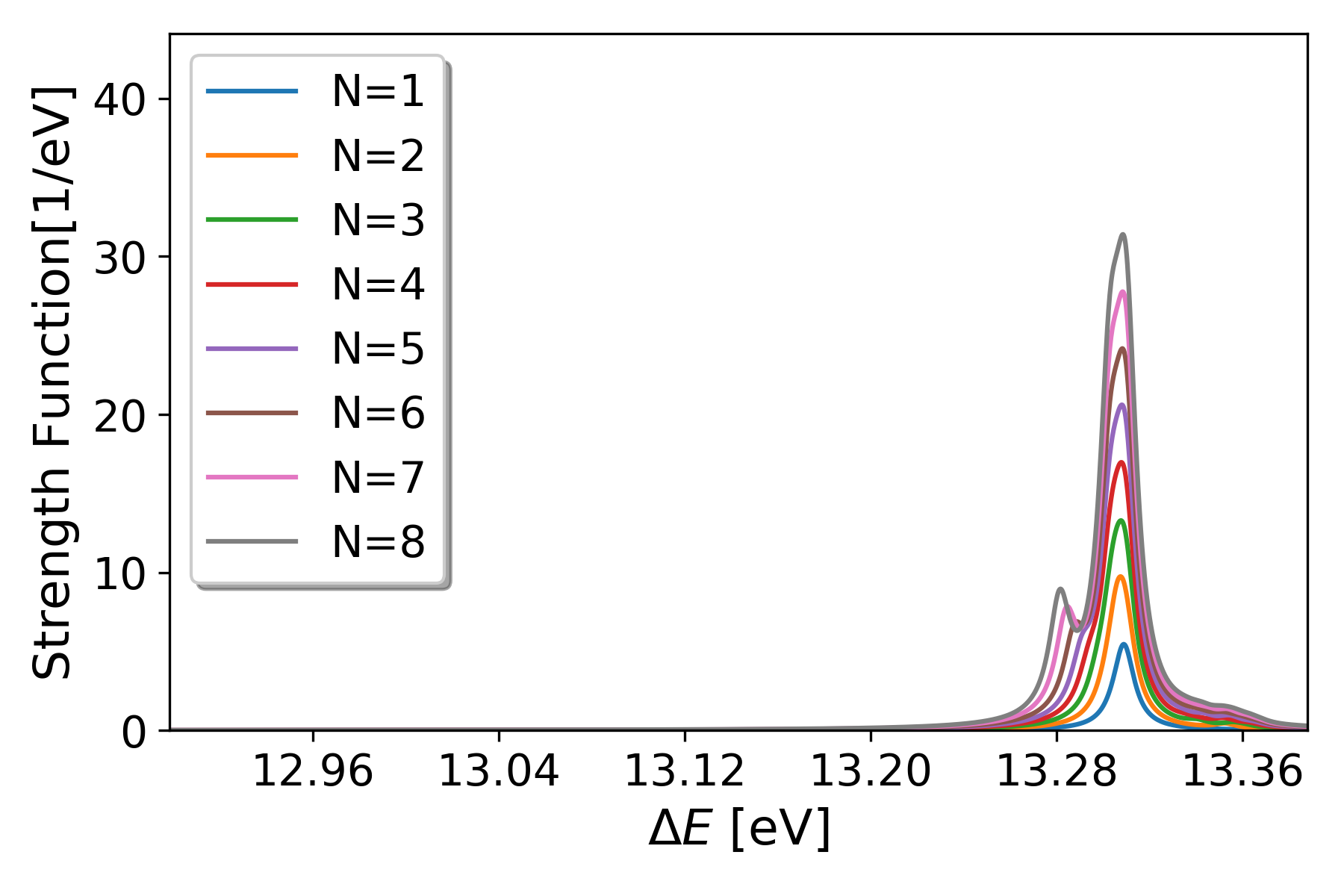}
    \label{fig:}
\end{subfigure}
\begin{subfigure}{.8\textwidth}
\centering
    \includegraphics[width=0.52\linewidth]{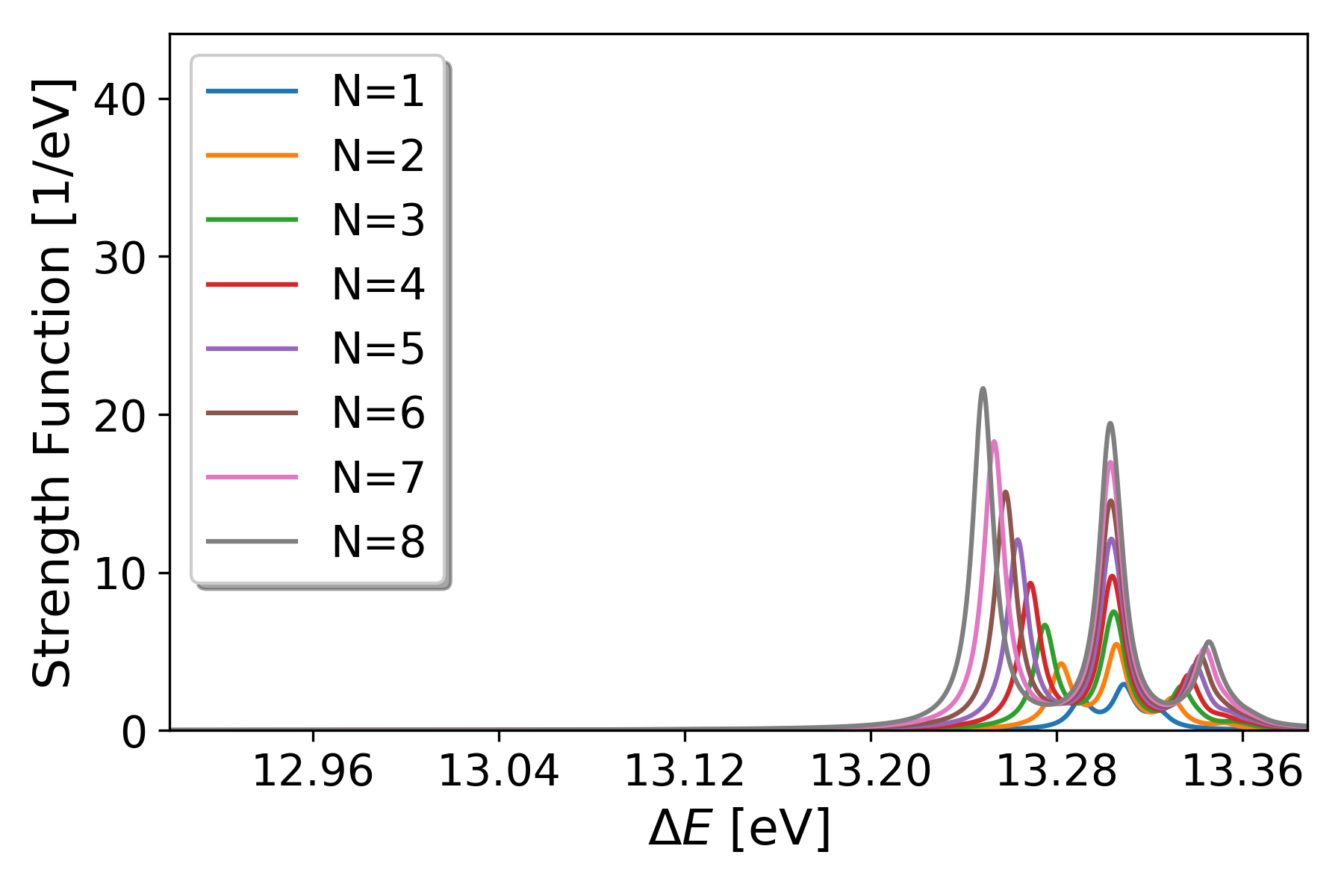}
    \label{fig:}
\end{subfigure}
\caption{Nitrogen dimer chain of variable size $N\in \{1..9\}$ without perturbed dimer. From top to bottom: $\boldsymbol{\lambda}=0$, $\boldsymbol{\lambda}\parallel \bold{e}_x$, $\boldsymbol{\lambda}\parallel \bold{e}_y$ and $\boldsymbol{\lambda}\parallel \bold{e}_z$.
}
\end{figure}

\begin{figure}[H]
\begin{subfigure}{.8\textwidth}
\centering
    \includegraphics[width=00.52\linewidth]{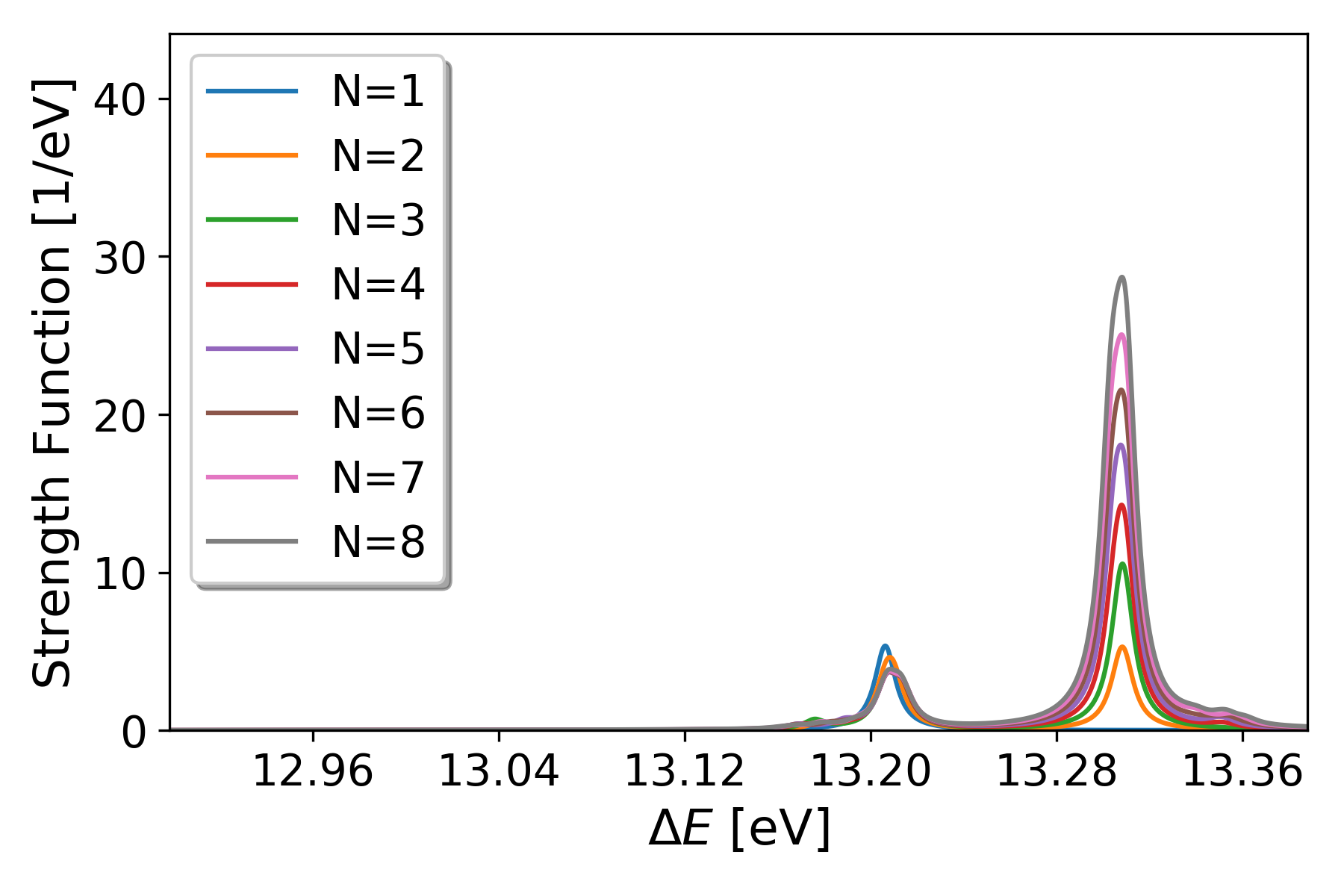}
    \label{fig:}
\end{subfigure}
\begin{subfigure}{.8\textwidth}
\centering
    \includegraphics[width=0.52\linewidth]{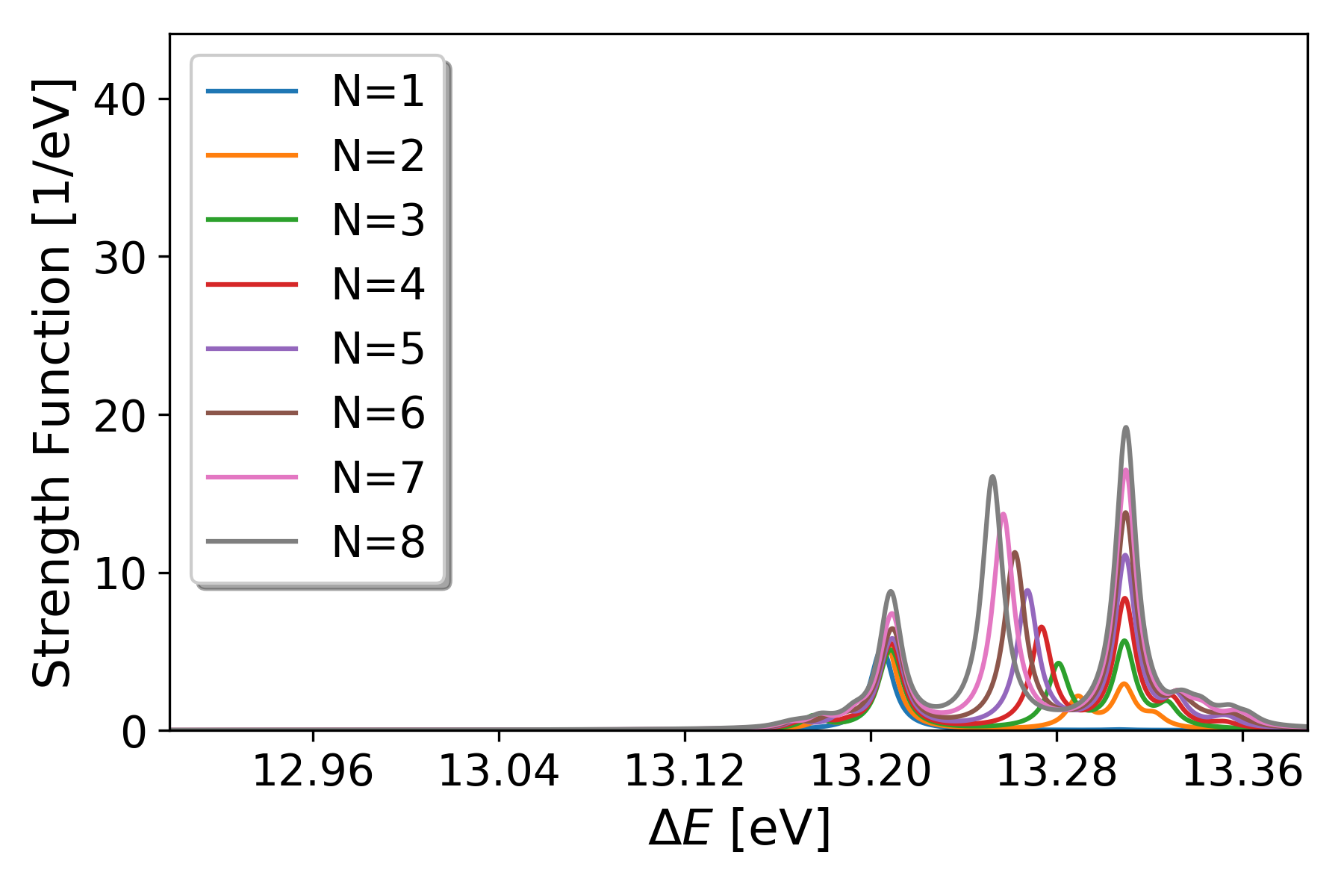}
    \label{fig:}
\end{subfigure}
\begin{subfigure}{.8\textwidth}
\centering
    \includegraphics[width=0.52\linewidth]{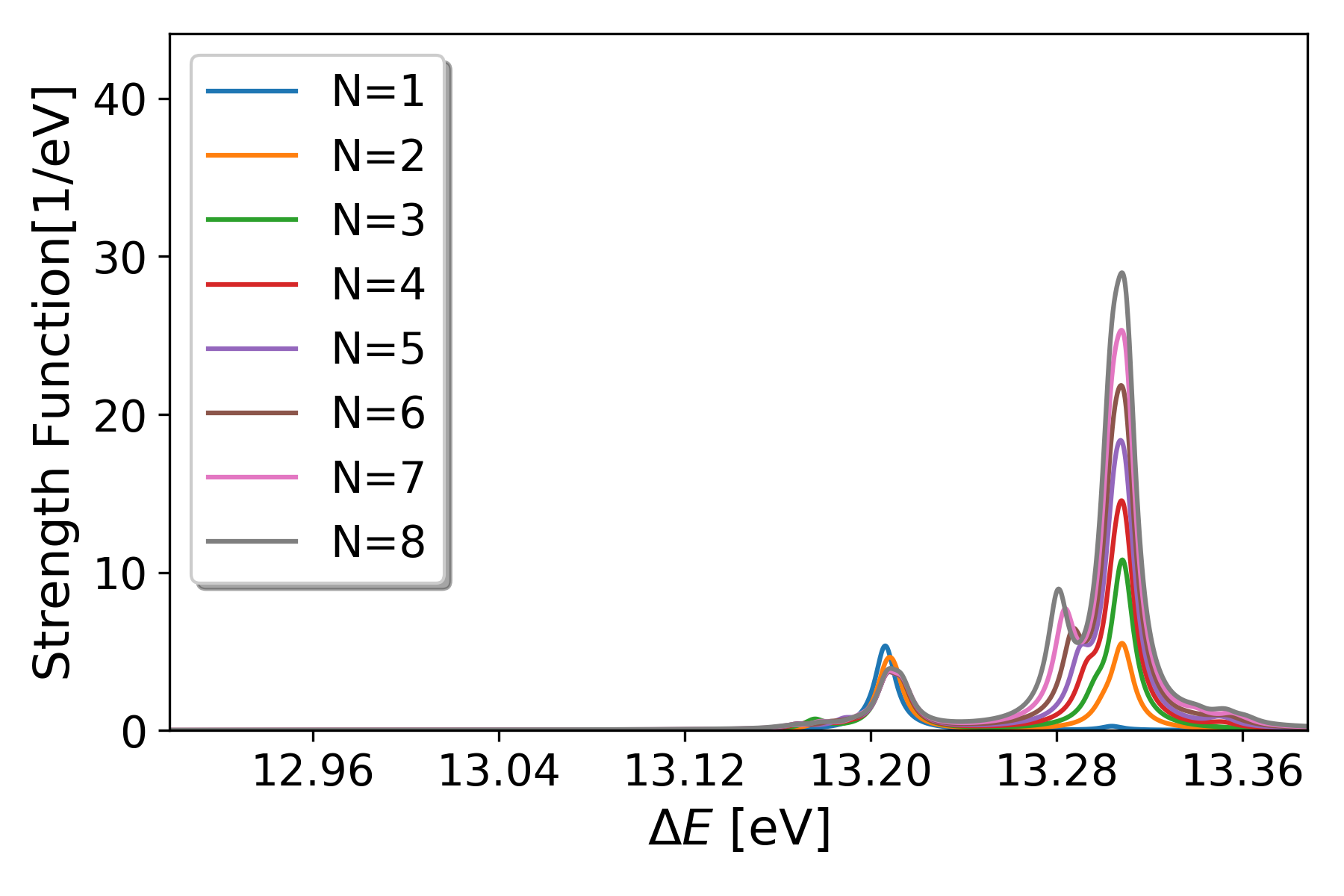}
    \label{fig:}
\end{subfigure}
\begin{subfigure}{.8\textwidth}
\centering
    \includegraphics[width=0.52\linewidth]{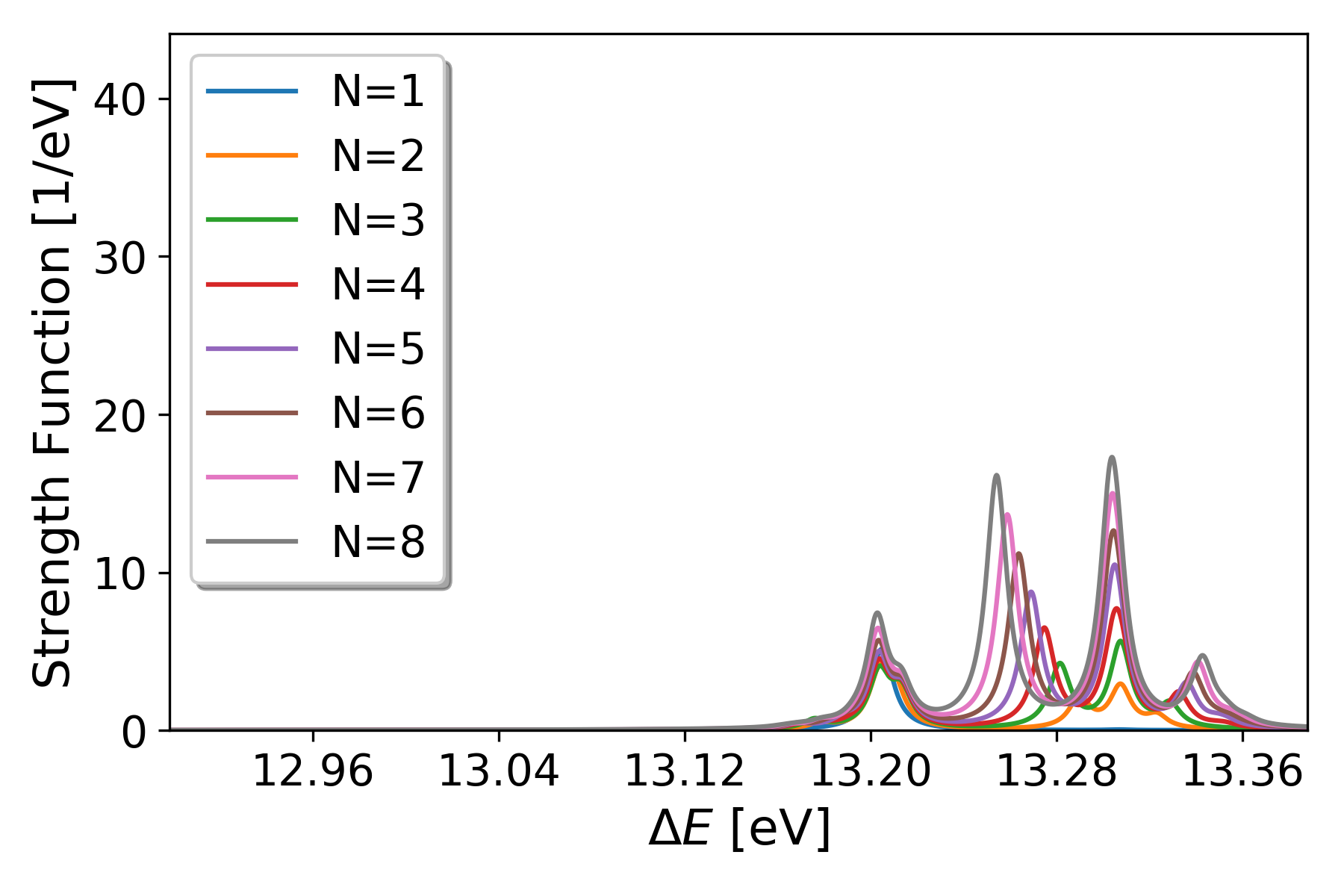}
    \label{fig:}
\end{subfigure}
\caption{Nitrogen dimer chain of variable size $N\in \{1..9\}$ with one perturbed dimer. From top to bottom: $\boldsymbol{\lambda}=0$, $\boldsymbol{\lambda}\parallel \bold{e}_x$, $\boldsymbol{\lambda}\parallel \bold{e}_y$ and $\boldsymbol{\lambda}\parallel \bold{e}_z$.
}
\end{figure}

\begin{figure}[H]
\begin{subfigure}{1\textwidth}
\centering
    \includegraphics[width=1\linewidth]{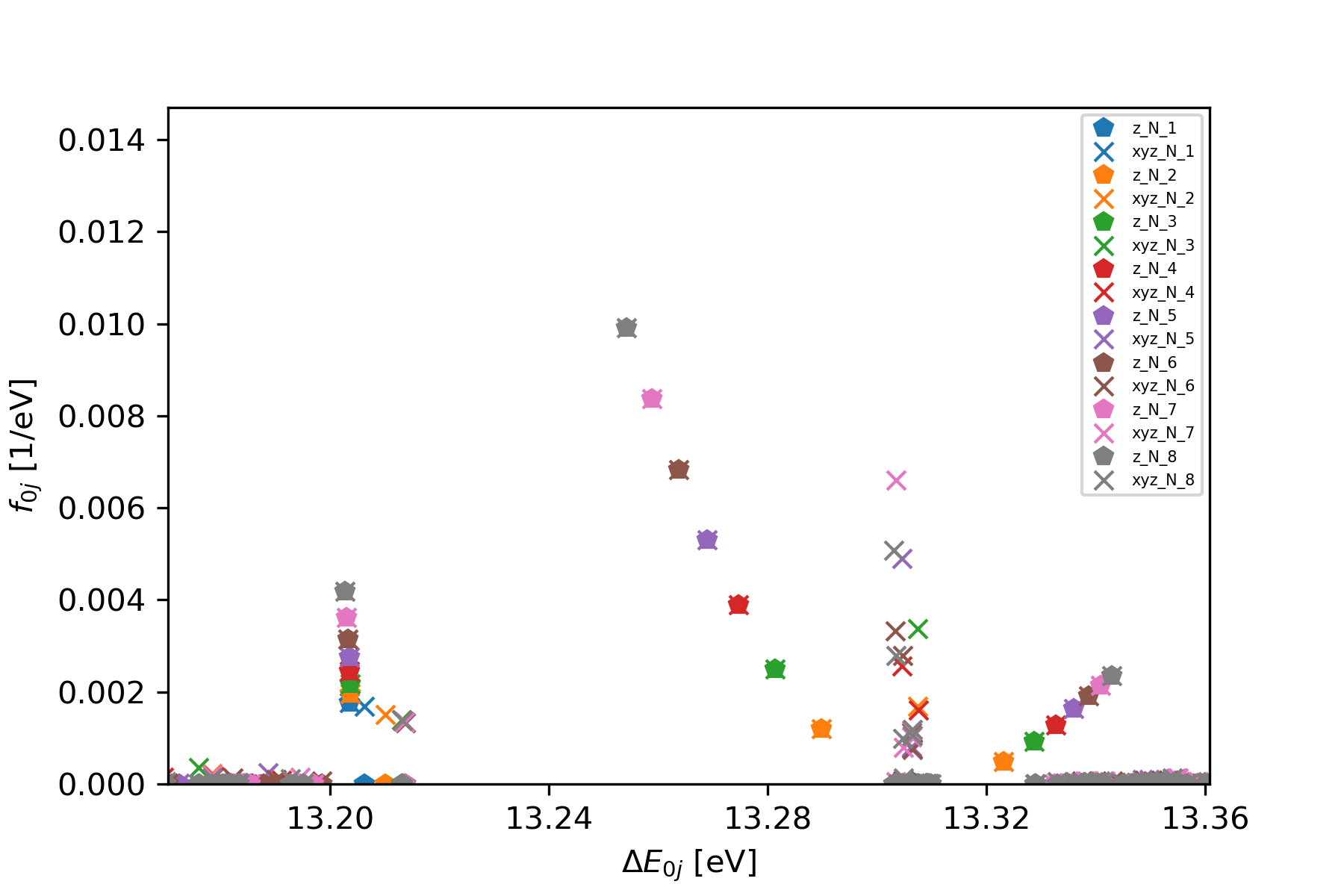}
\end{subfigure}
    \caption{Oscillator strengths for perturbed dimer chain of variable size, with cavity oriented along $z$-axis in coupling regime \RomanNumeralCaps{1}. Bold symbols indicate only contributions from transition dipolements along $z$ whereas crosses account equally weighted for all three transition dipole moments along $x,y,z$. The later acts as a basis for the standard Lorentz-broadened spectra.}
\label{fig:middle_p_scal}
\end{figure}

\subsubsection{Local Properties}

\begin{figure}[H]
\begin{subfigure}{1\textwidth}
\centering
    \includegraphics[trim=60 0 0 0, clip, width=0.9\linewidth]{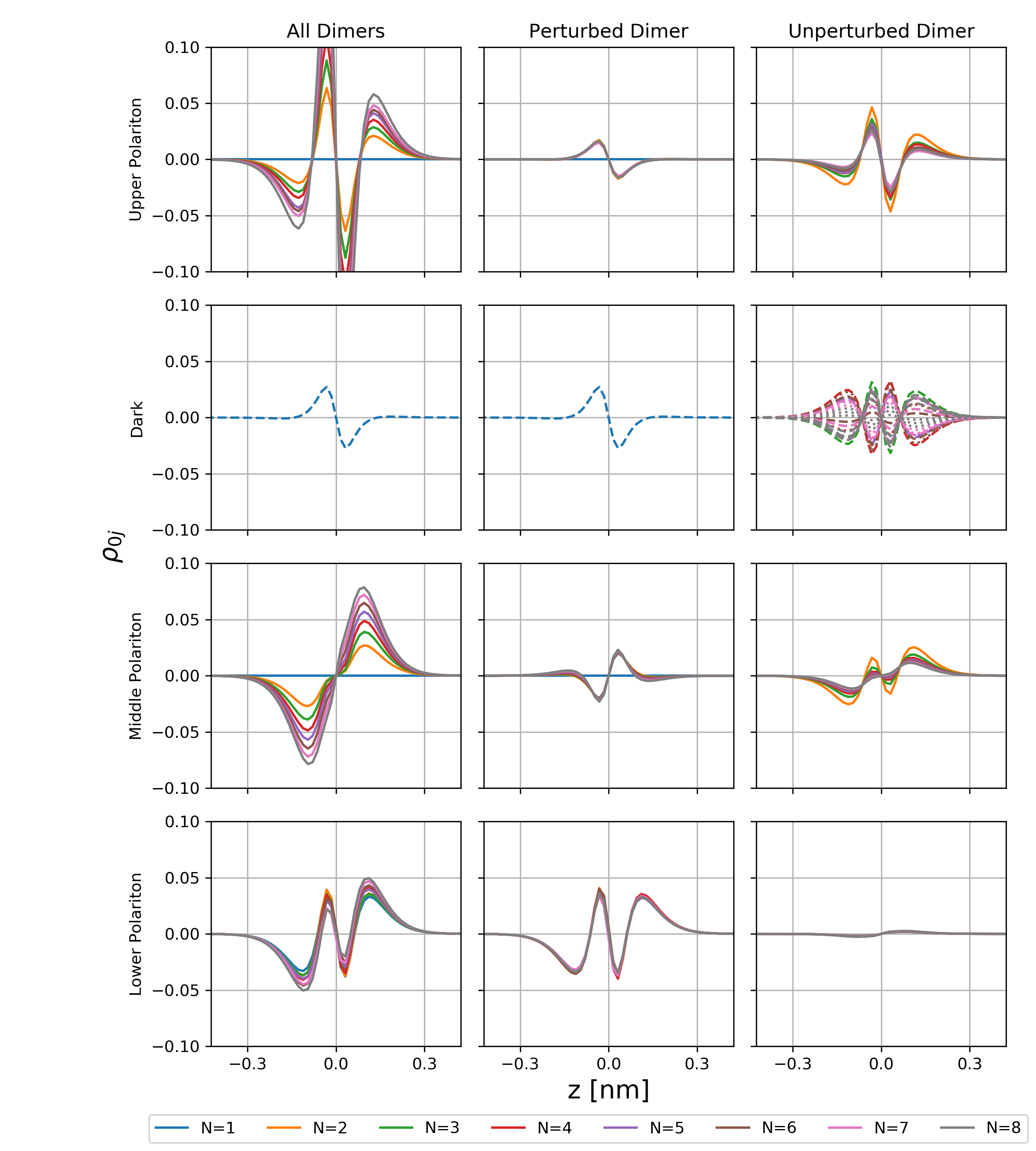}
\end{subfigure}
    \caption{Globally (left column) and locally resolved transition densities projected onto the $z$-axis for different chain lengths $N$ in coupling regime \RomanNumeralCaps{1}. For each of the four energy windows (rows), integrated quantities are displayed, except for the dark states. The integration cleans the data and contributes only very little to the overall results. }
\label{fig:middle_p_scal}
\end{figure}

\begin{figure}[H]
\begin{subfigure}{1\textwidth}
\centering
    \includegraphics[trim=60 0 0 0, clip, width=1\linewidth]{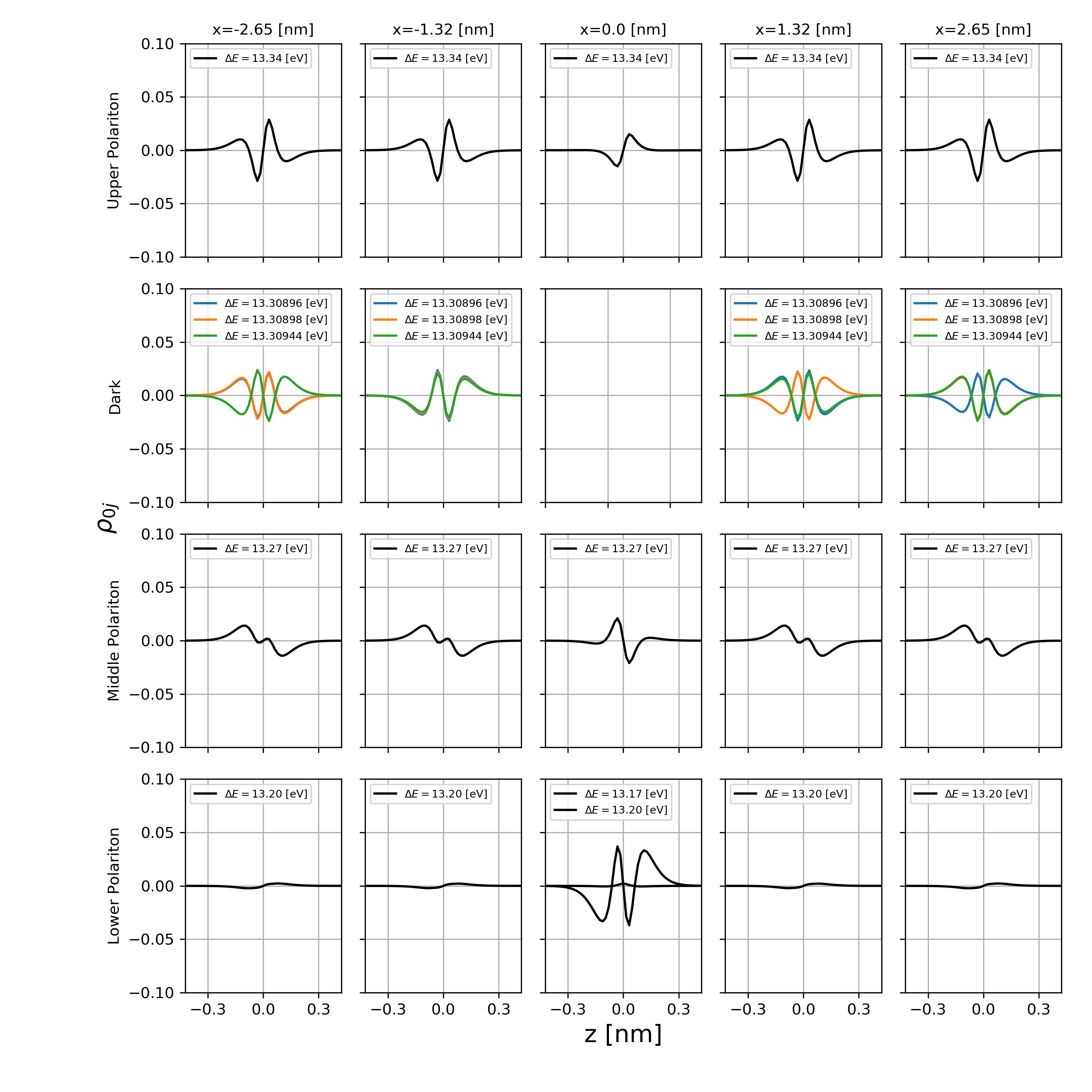}
\end{subfigure}
    \caption{Locally resolved transition densities of each dimer projected onto the $z$-axis for $N=5$ and coupling regime \RomanNumeralCaps{1}. For each of the four energy windows (rows), integrated quantities are displayed, except for the three emerging dark states.}
\label{fig:td_loc_Nd_5}
\end{figure}

\section{Simulation Results for Coupling Regime \RomanNumeralCaps{2} ($\lambda=0.005$)}
\subsection{Absorption Spectra for Cavity in Resonance with Unperturbed Dimers}

\begin{figure}[H]
\begin{subfigure}{.8\textwidth}
\centering
    \includegraphics[width=0.52\linewidth]{figures_SI/Spec_min_D_para_y_ddist_25b_casida_N.png}
    \label{fig:}
\end{subfigure}
\begin{subfigure}{.8\textwidth}
\centering
    \includegraphics[width=0.52\linewidth]{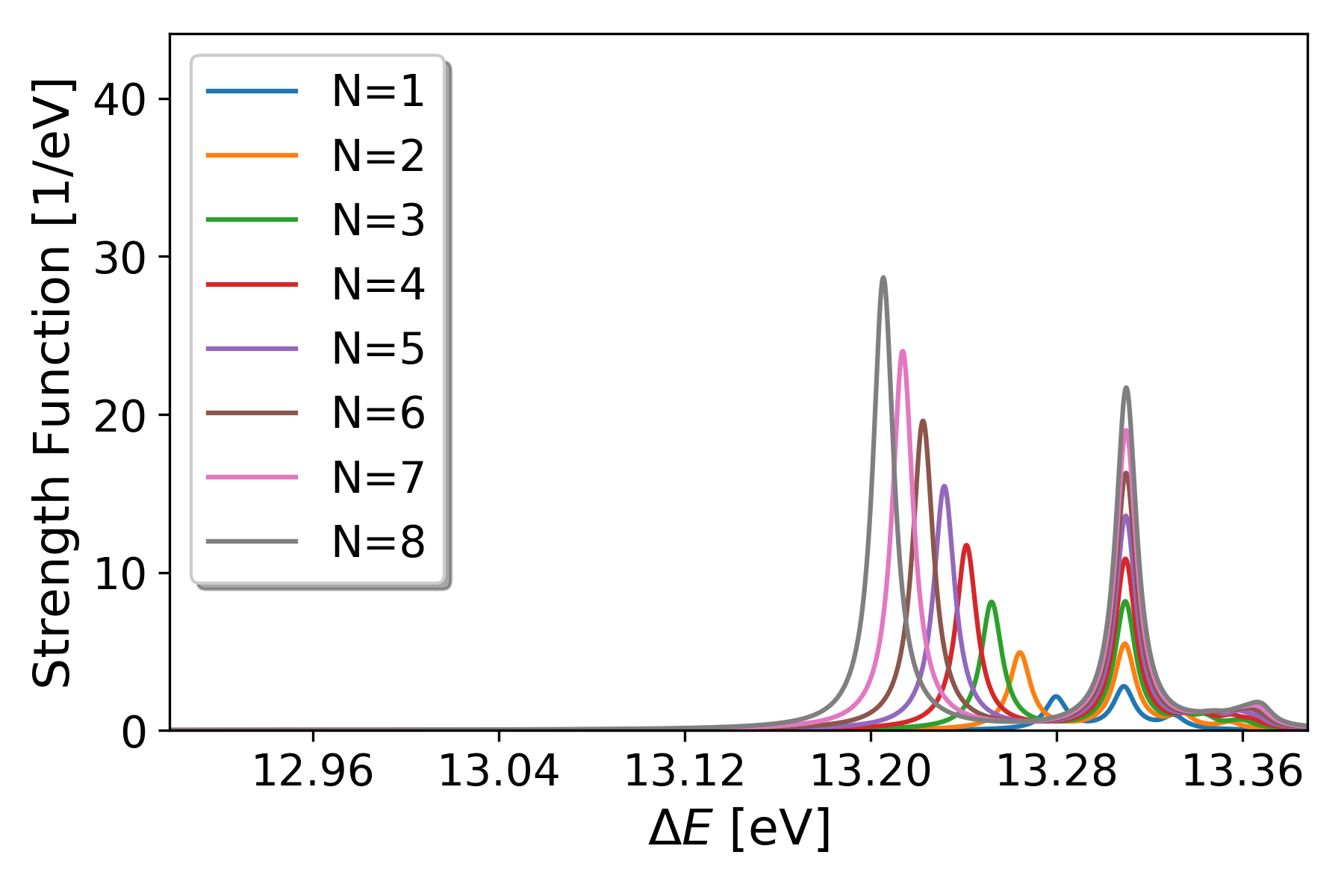}
    \label{fig:}
\end{subfigure}
\begin{subfigure}{.8\textwidth}
\centering
    \includegraphics[width=0.52\linewidth]{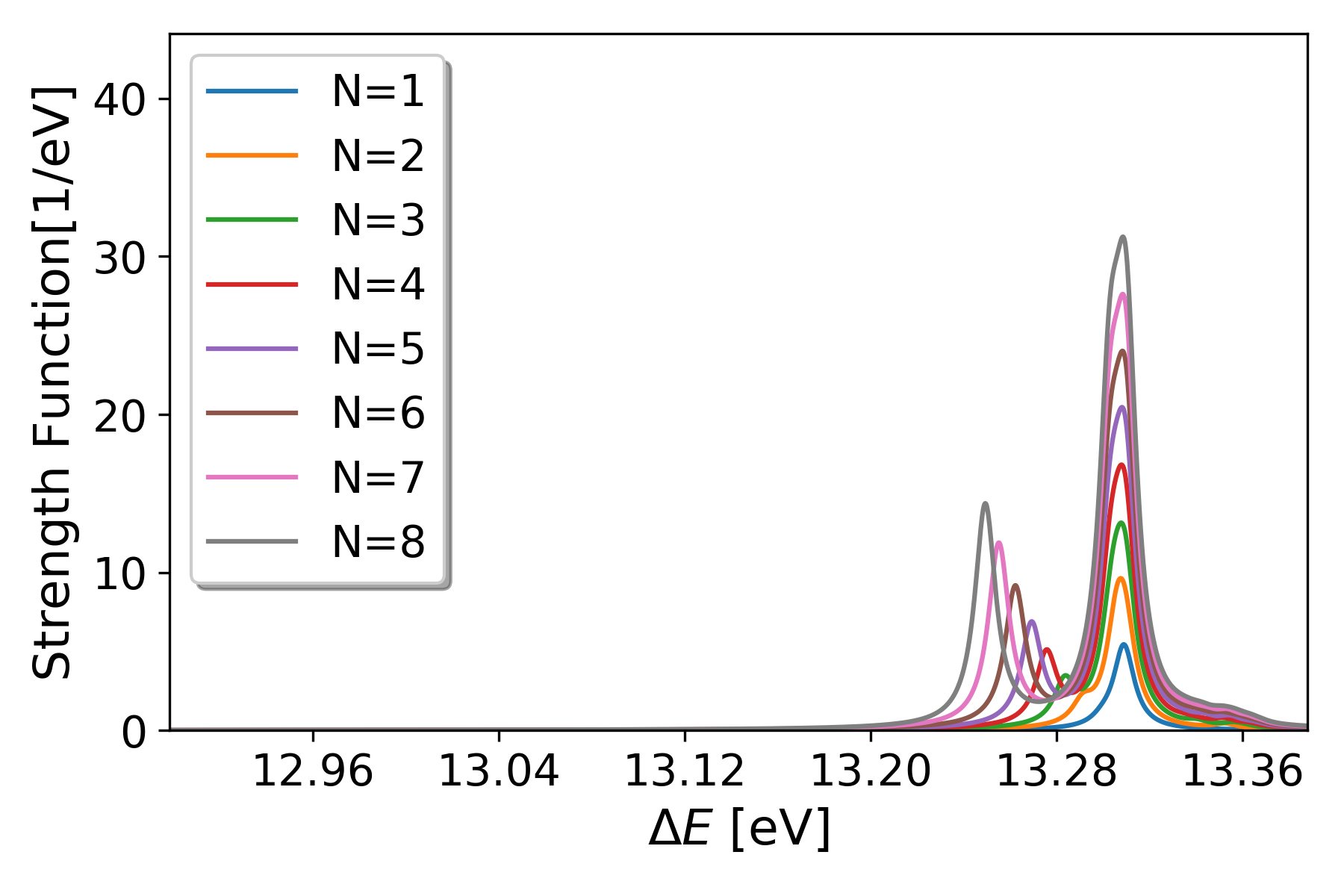}
    \label{fig:}
\end{subfigure}
\begin{subfigure}{.8\textwidth}
\centering
    \includegraphics[width=0.52\linewidth]{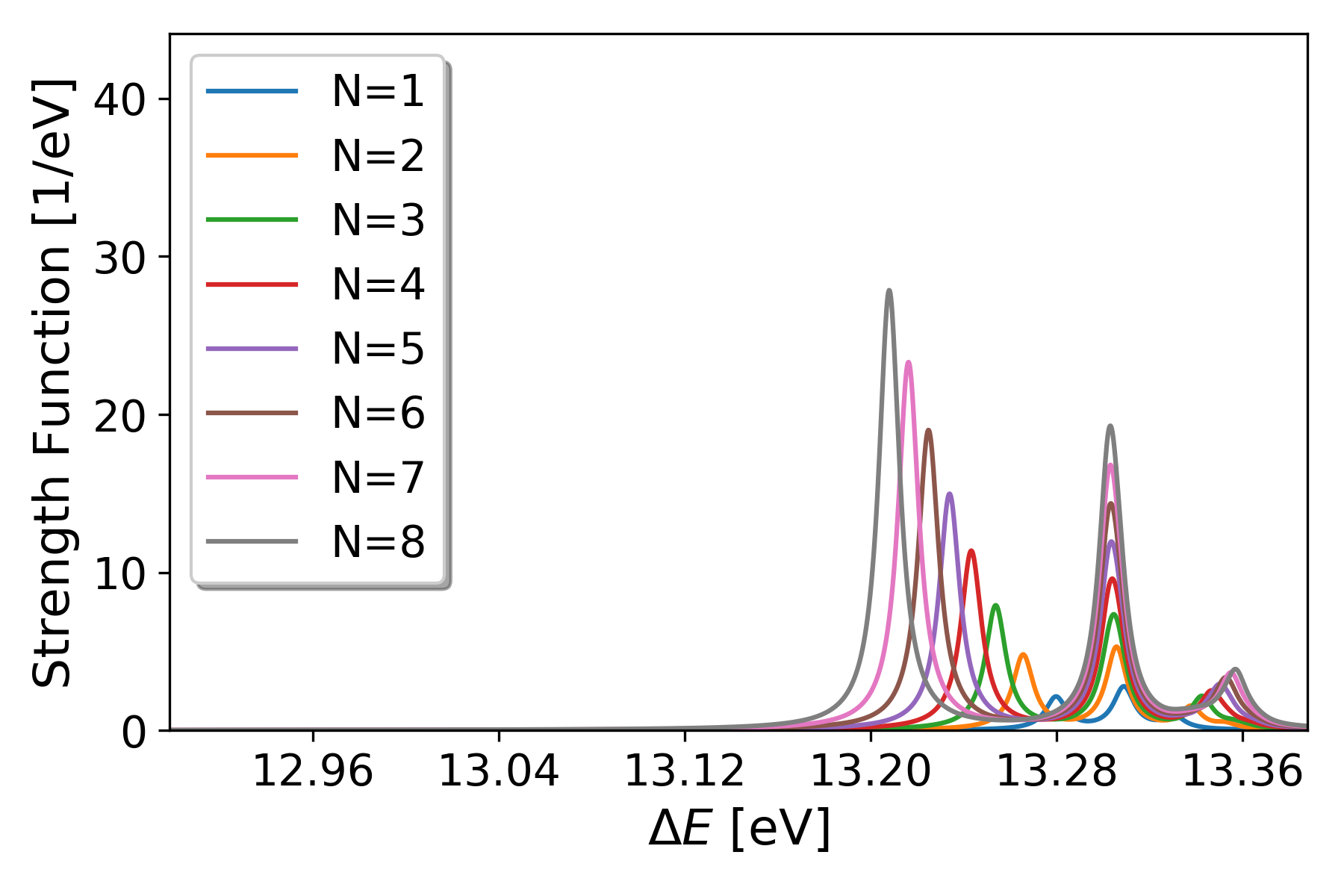}
    \label{fig:}
\end{subfigure}
\caption{Nitrogen dimer chain of variable size $N\in \{1..9\}$ without perturbed dimer. From top to bottom: $\boldsymbol{\lambda}=0$, $\boldsymbol{\lambda}\parallel \bold{e}_x$, $\boldsymbol{\lambda}\parallel \bold{e}_y$ and $\boldsymbol{\lambda}\parallel \bold{e}_z$.
}
\end{figure}

\begin{figure}[H]
\begin{subfigure}{.8\textwidth}
\centering
    \includegraphics[width=0.52\linewidth]{figures_SI/Spec_pert_D_para_y_ddist_25b_casida_N.png}
    \label{fig:}
\end{subfigure}
\begin{subfigure}{.8\textwidth}
\centering
    \includegraphics[width=0.52\linewidth]{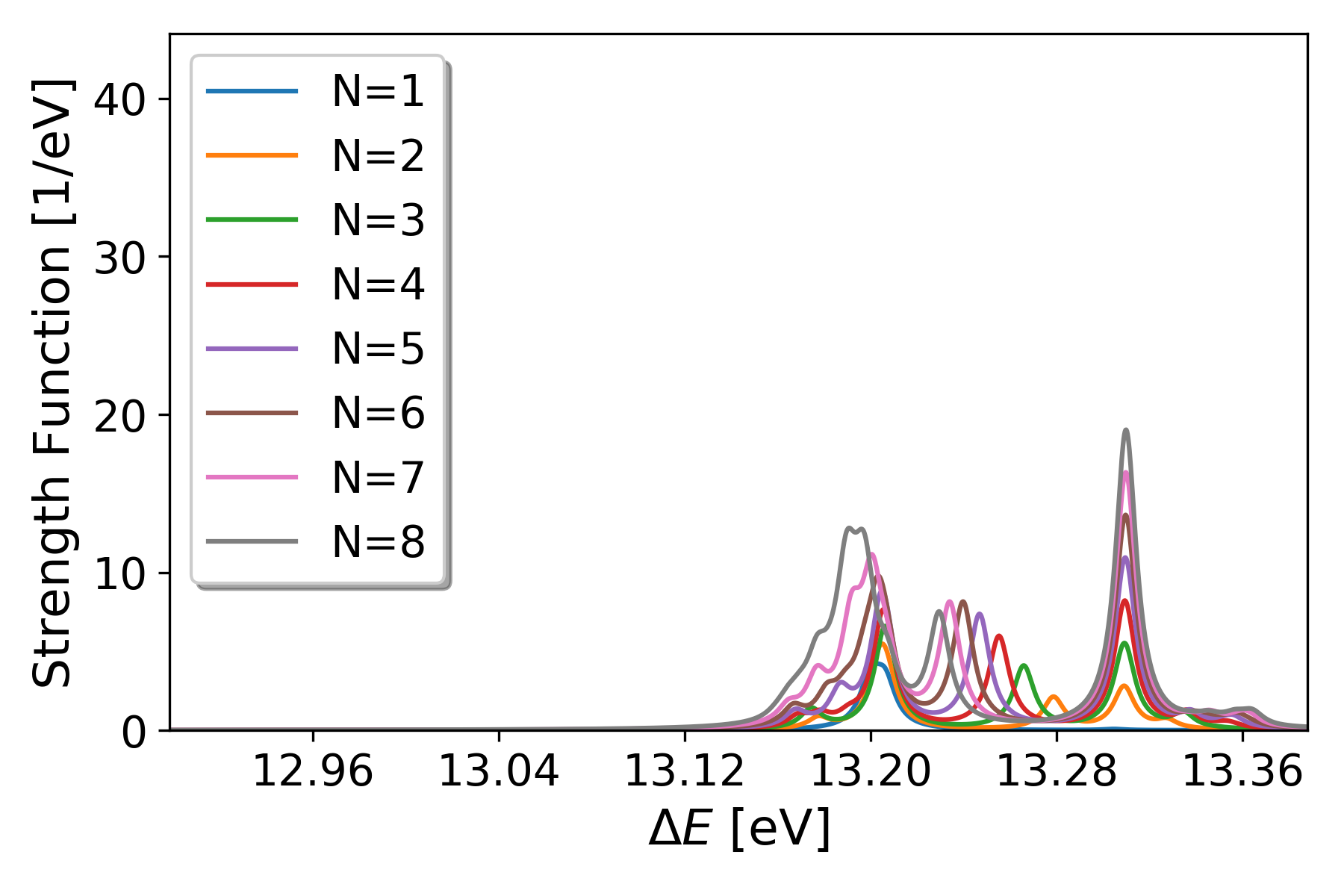}
    \label{fig:}
\end{subfigure}
\begin{subfigure}{.8\textwidth}
\centering
    \includegraphics[width=0.52\linewidth]{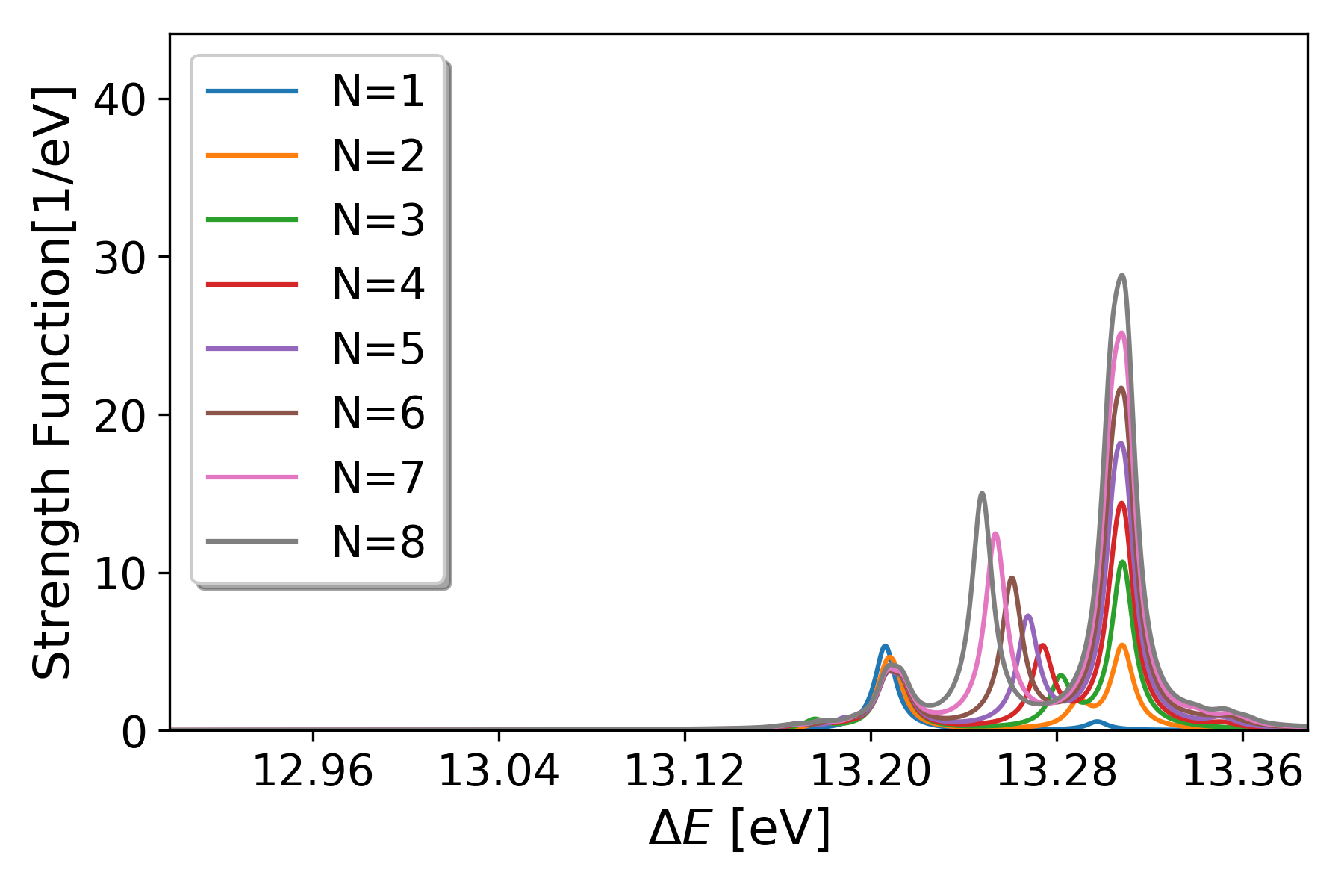}
    \label{fig:}
\end{subfigure}
\begin{subfigure}{.8\textwidth}
\centering
    \includegraphics[width=0.52\linewidth]{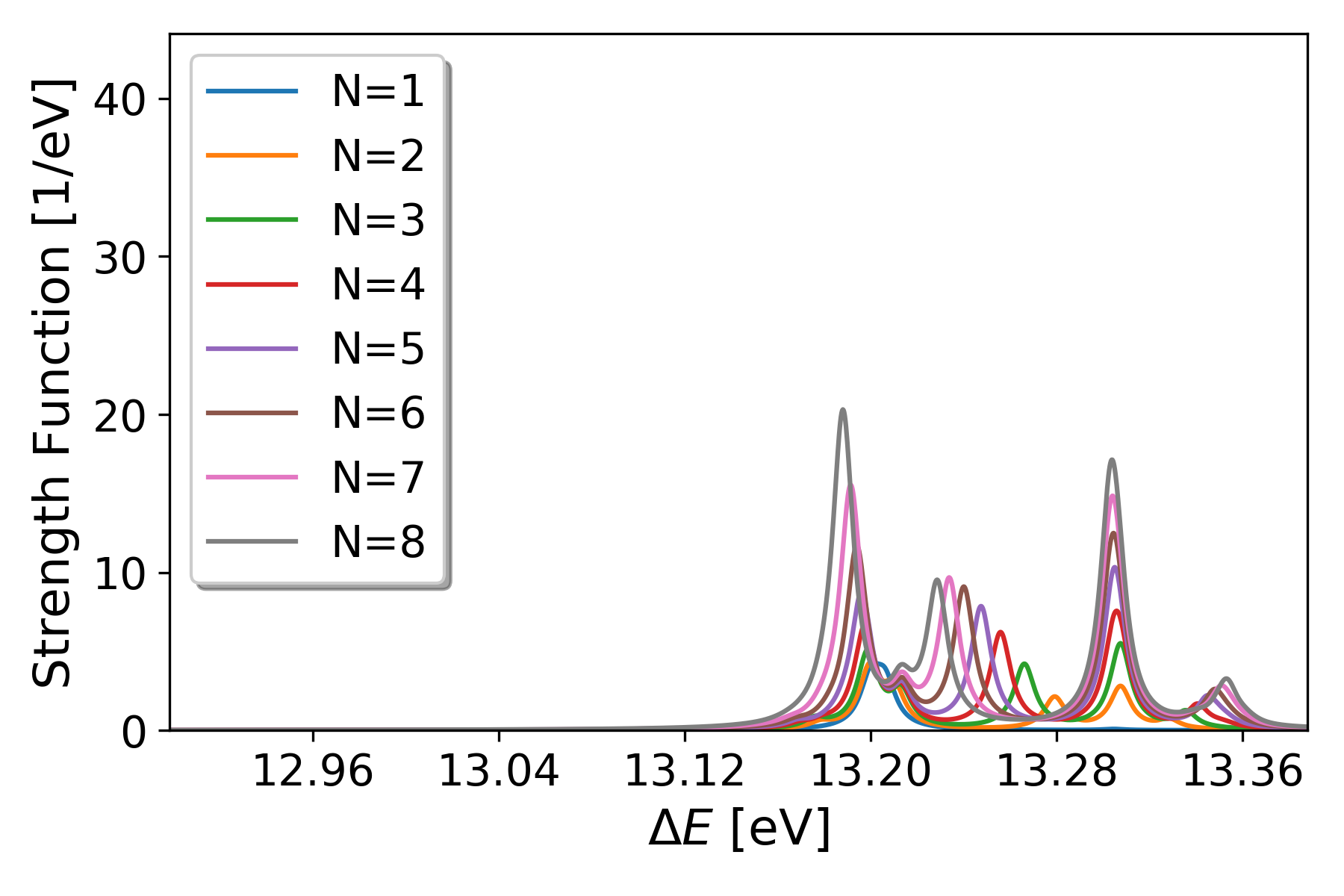}
    \label{fig:}
\end{subfigure}
\caption{Nitrogen dimer chain of variable size $N\in \{1..9\}$ with one perturbed dimer. From top to bottom: $\boldsymbol{\lambda}=0$, $\boldsymbol{\lambda}\parallel \bold{e}_x$, $\boldsymbol{\lambda}\parallel \bold{e}_y$ and $\boldsymbol{\lambda}\parallel \bold{e}_z$.
}
\end{figure}

\begin{figure}[H]
\begin{subfigure}{1\textwidth}
\centering
    \includegraphics[width=1\linewidth]{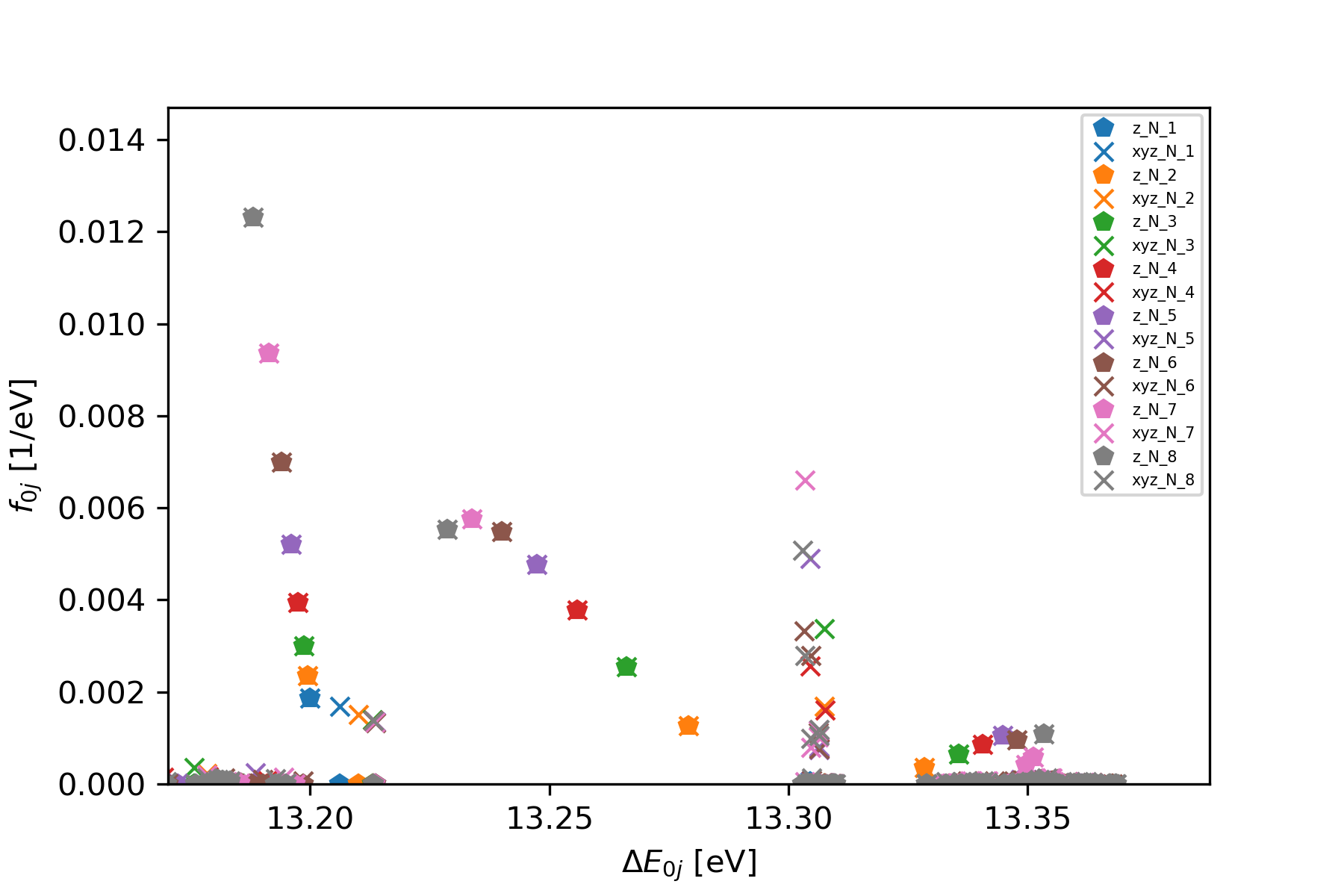}
\end{subfigure}
    \caption{Oscillator strengths for perturbed dimer chain of variable size, with cavity oriented along $z$-axis in coupling regime \RomanNumeralCaps{2}. Bold symbols indicate only contributions from transition dipolements along $z$ whereas crosses account equally weighted for all three transition dipole moments along $x,y,z$. The later acts as a basis for the previously shown Lorentz-broadened spectra. Notice the splitting in the upper polaritonic branch for $N=7$ (labeled $\mathrm{z\_N\_7}$ in purple), which is caused by a sign change of the $N-1$ unperturbed transition dipole moments. Whether or not real systems can be prepared to enter this novel regime, will be topic of future investigations.}
\label{fig:middle_p_scal}
\end{figure}

\begin{figure}[H]
\begin{subfigure}{1\textwidth}
\centering
    \includegraphics[width=1\linewidth]{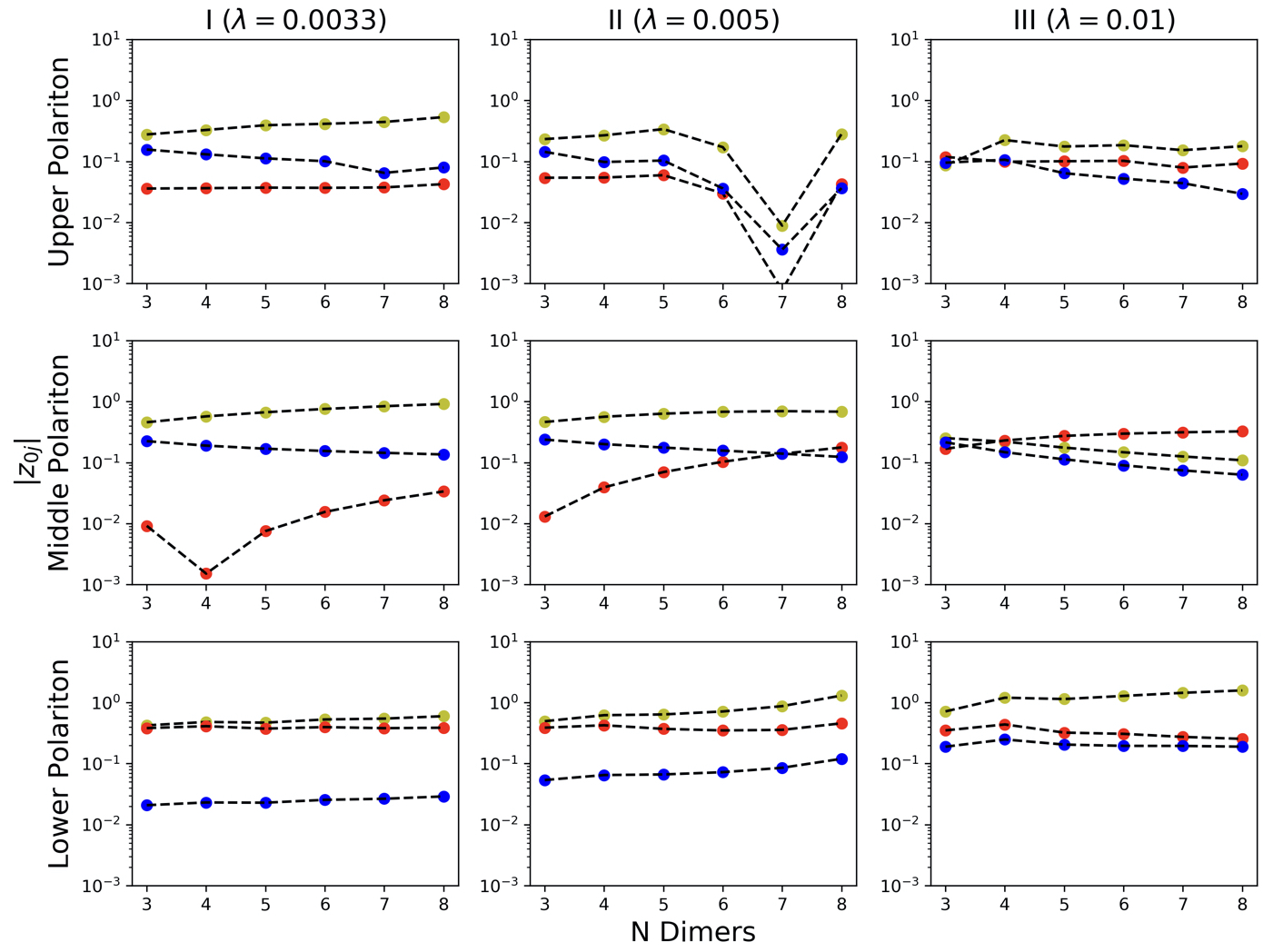}
\end{subfigure}
    \caption{Global (yellow) and local transition dipole moment scaling with respect to different chain lengths for the perturbed (red) and unperturbed (blue) dimers. Special cases $N=\{1,2\}$ are excluded for the sake of clarity (either does not include an unperturbed dimer or no dark states can form). }
\label{fig:dipole_scale}
\end{figure}

\subsubsection{Mimic collectively induced modifications of the impurity with a suitable choice of $\lambda$}

\begin{figure}[H]
\begin{subfigure}{1\textwidth}
\centering
    \includegraphics[width=0.7\linewidth]{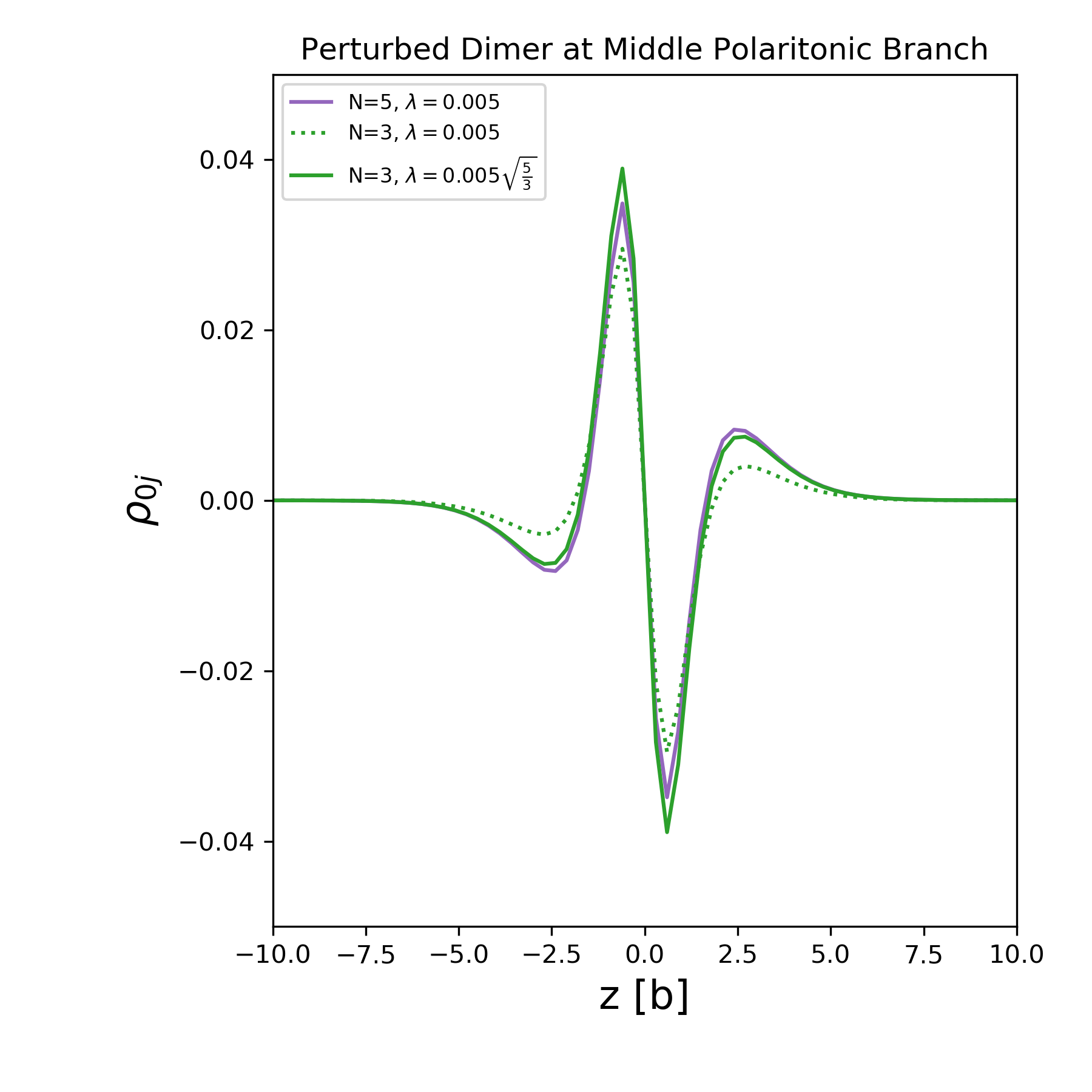}
\end{subfigure}
    \caption{Local transition density for the perturbed dimer in the middle polaritonic branch for coupling regime \RomanNumeralCaps{2}. The scaling up of $\lambda$ for $N=3$ allows to mimic locally imposed  effects on the perturbed dimer, which arise from collective coupling in larger systems (here $N=5$). This opens the door for efficient \textit{ab initio} simulation methods in polaritonic chemistry.  }
\label{fig:middle_p_scal}
\end{figure}

\begin{figure}[H]
\begin{subfigure}{1\textwidth}
\centering
    \includegraphics[trim=60 0 0 0, clip, width=1\linewidth]{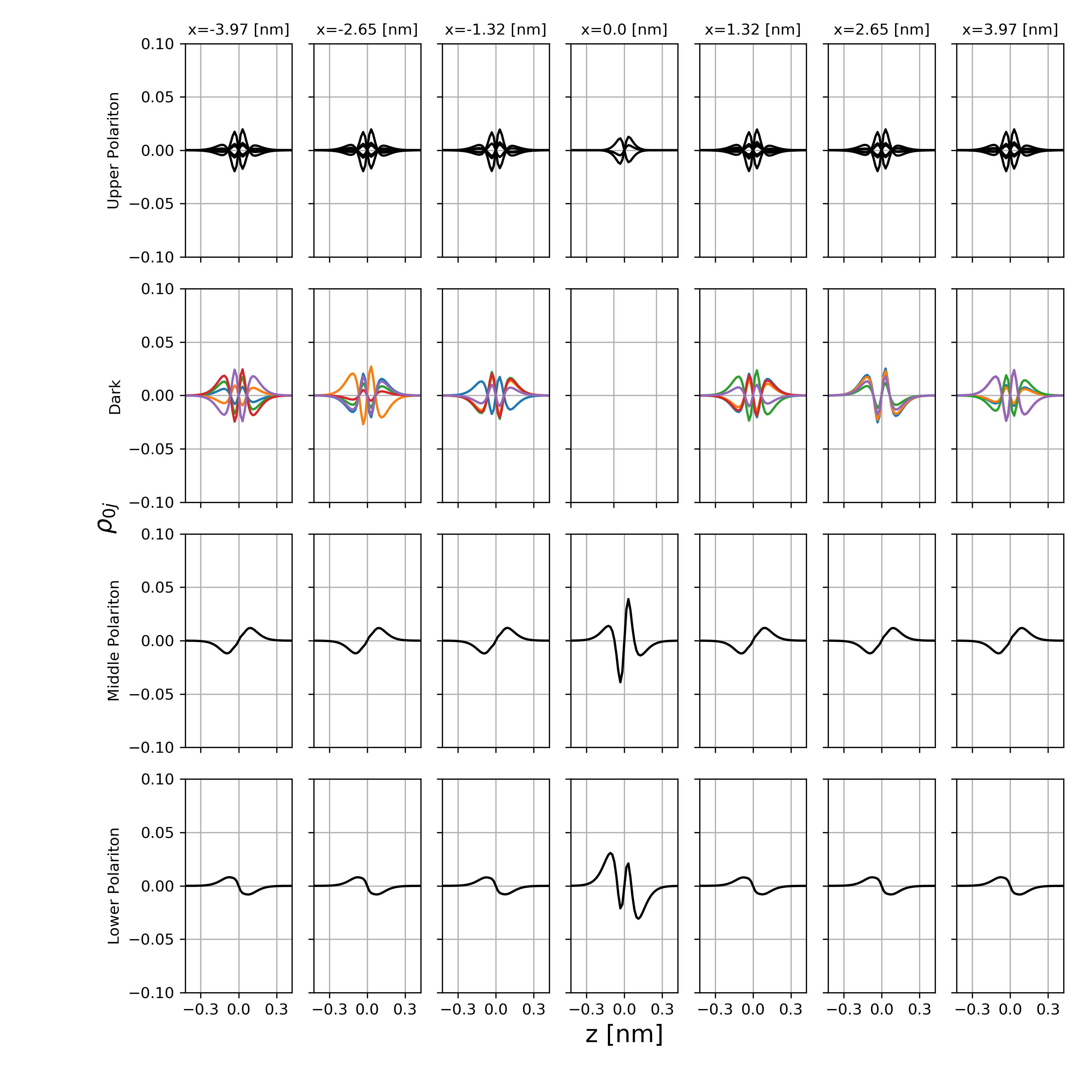}
\end{subfigure}
    \caption{Locally resolved transition densities of each dimer projected onto the $z$-axis for $N=7$ and coupling regime \RomanNumeralCaps{2}. Cavity tuned to unperturbed frequency $\omega_u$ and $\lambda=0.005$. For each of the four energy windows (rows), integrated quantities are displayed, except for the emerging dark states (in color). Notice that the different dark state patterns have significantly different local transition densities.}
\label{fig:td_loc_Nd_7}
\end{figure}

\subsection{Absorption Spectra for Cavity in Resonance with Perturbation}

\begin{figure}[H]
\begin{subfigure}{.8\textwidth}
\centering
    \includegraphics[width=0.52\linewidth]{figures_SI/Spec_min_D_para_y_ddist_25b_casida_N.png}
    \label{fig:}
\end{subfigure}
\begin{subfigure}{.8\textwidth}
\centering
    \includegraphics[width=0.52\linewidth]{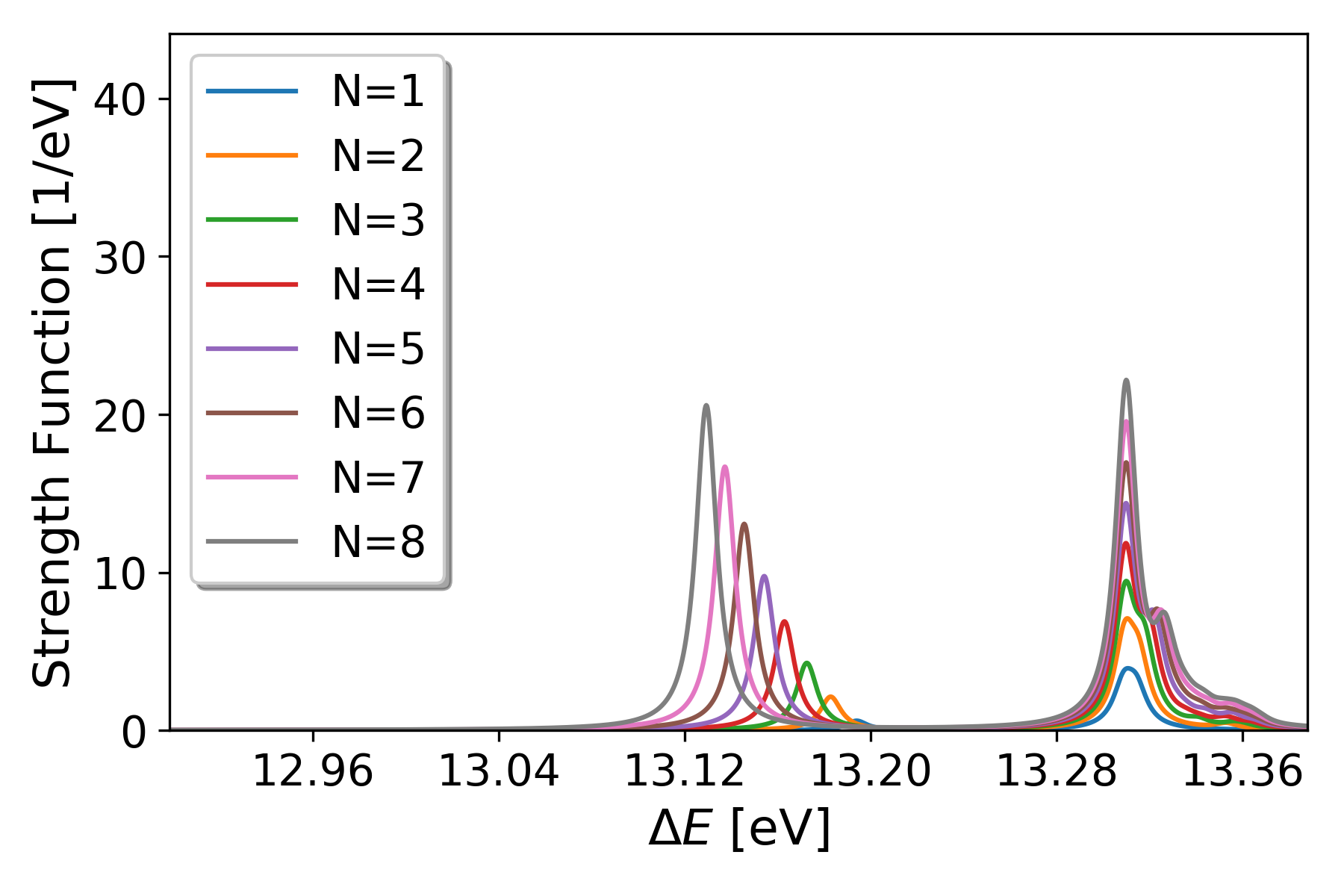}
    \label{fig:}
\end{subfigure}
\begin{subfigure}{.8\textwidth}
\centering
    \includegraphics[width=0.52\linewidth]{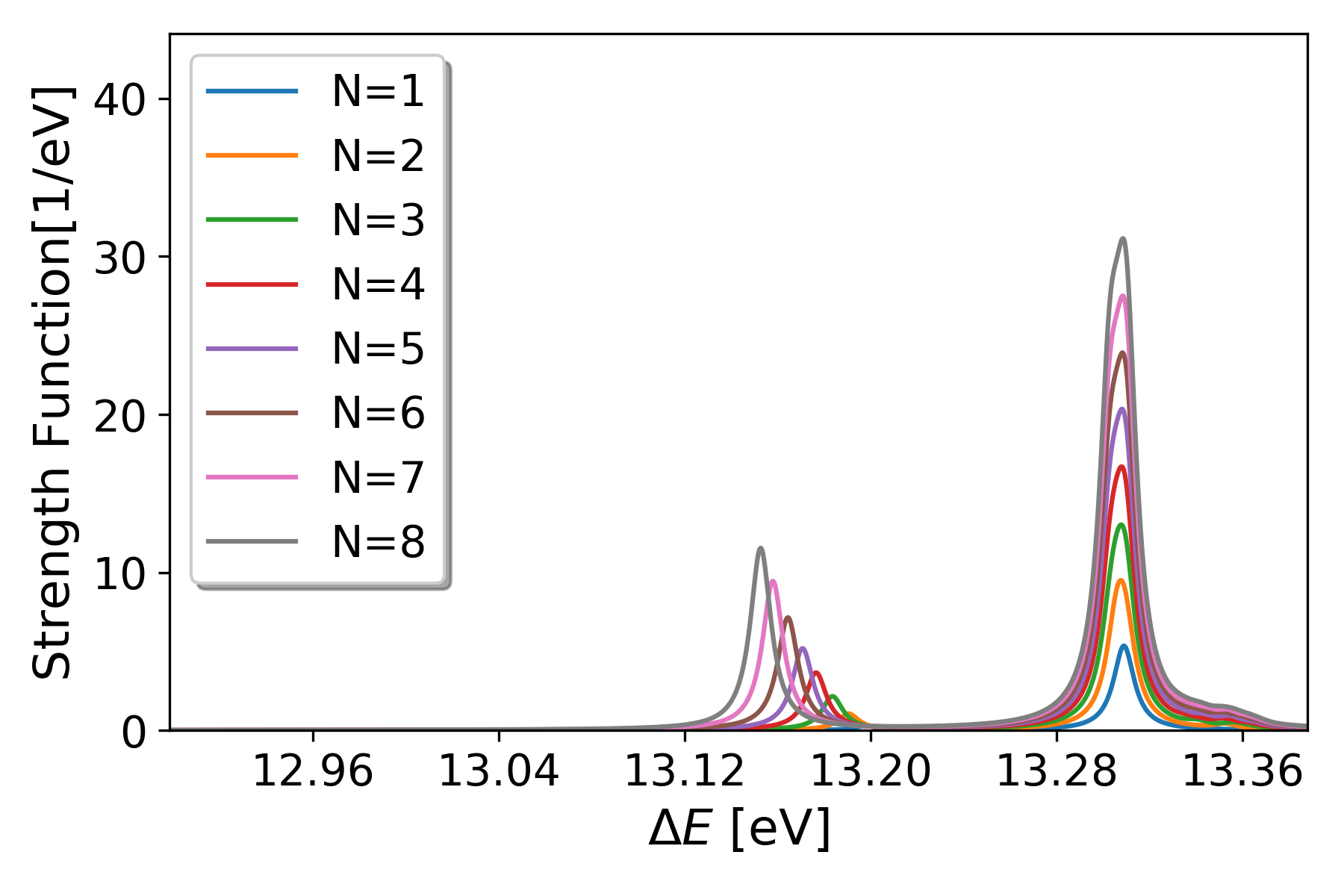}
    \label{fig:}
\end{subfigure}
\begin{subfigure}{.8\textwidth}
\centering
    \includegraphics[width=0.52\linewidth]{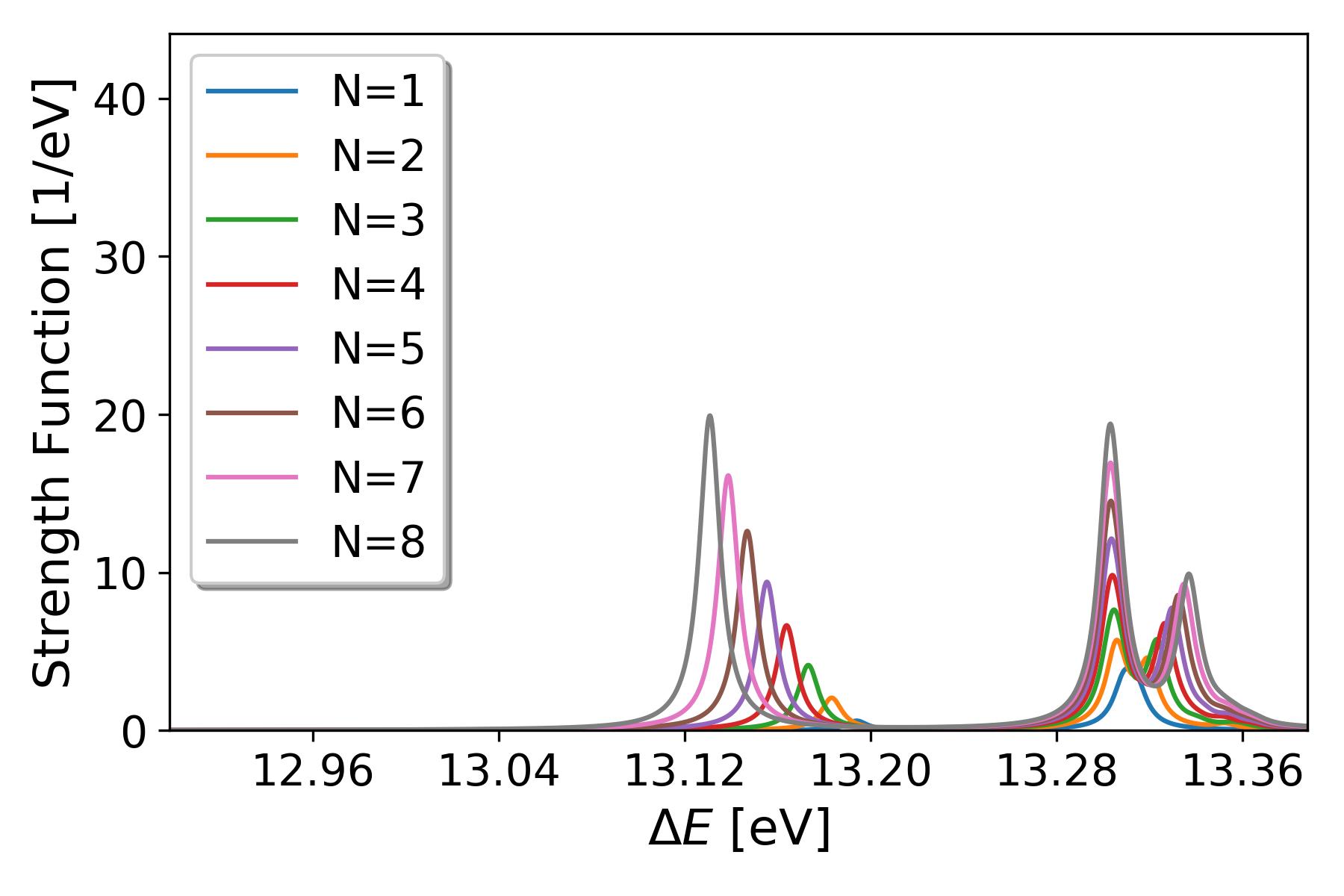}
    \label{fig:}
\end{subfigure}
\caption{Nitrogen dimer chain of variable size $N\in \{1..9\}$ without perturbed dimer. Cavity tuned to perturbed frequency $\omega_p$ (!) and $\lambda=0.005$. From top to bottom: $\boldsymbol{\lambda}=0$, $\boldsymbol{\lambda}\parallel \bold{e}_x$, $\boldsymbol{\lambda}\parallel \bold{e}_y$ and $\boldsymbol{\lambda}\parallel \bold{e}_z$.
}
\end{figure}

\begin{figure}[H]
\begin{subfigure}{.8\textwidth}
\centering
    \includegraphics[width=0.52\linewidth]{figures_SI/Spec_pert_D_para_y_ddist_25b_casida_N.png}
    \label{fig:}
\end{subfigure}
\begin{subfigure}{.8\textwidth}
\centering
    \includegraphics[width=0.52\linewidth]{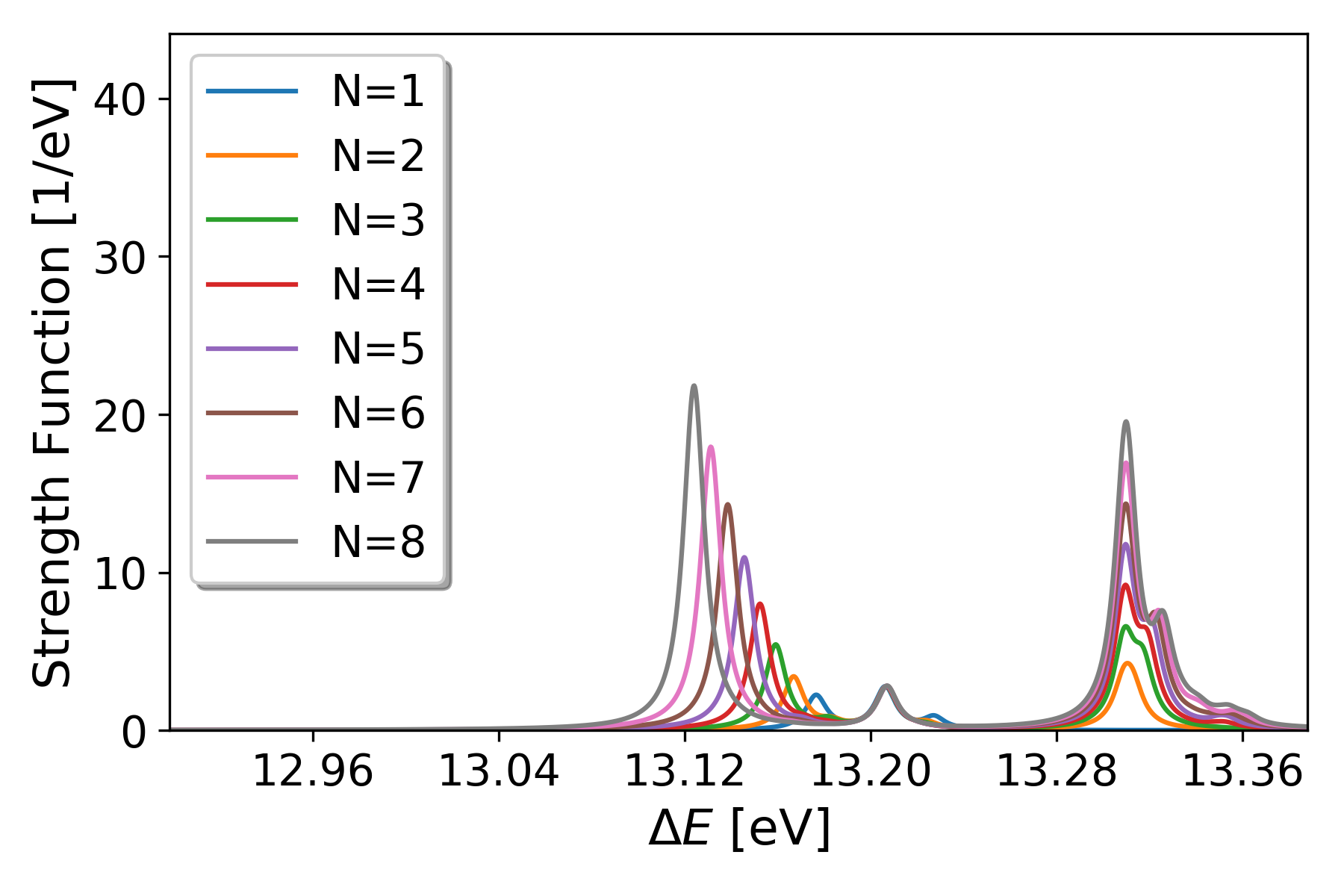}
    \label{fig:}
\end{subfigure}
\begin{subfigure}{.8\textwidth}
\centering
    \includegraphics[width=0.52\linewidth]{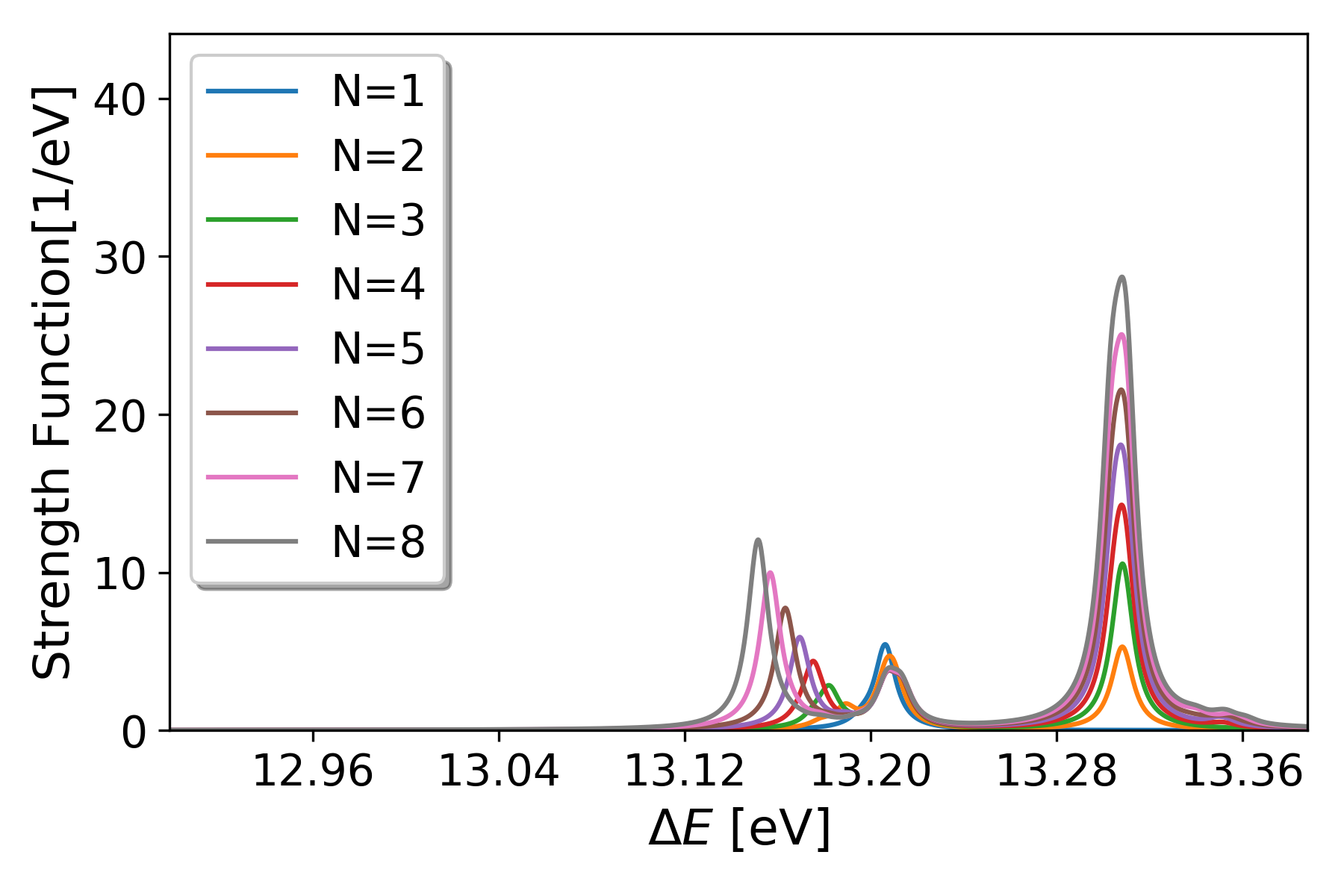}
    \label{fig:}
\end{subfigure}
\begin{subfigure}{.8\textwidth}
\centering
    \includegraphics[width=0.52\linewidth]{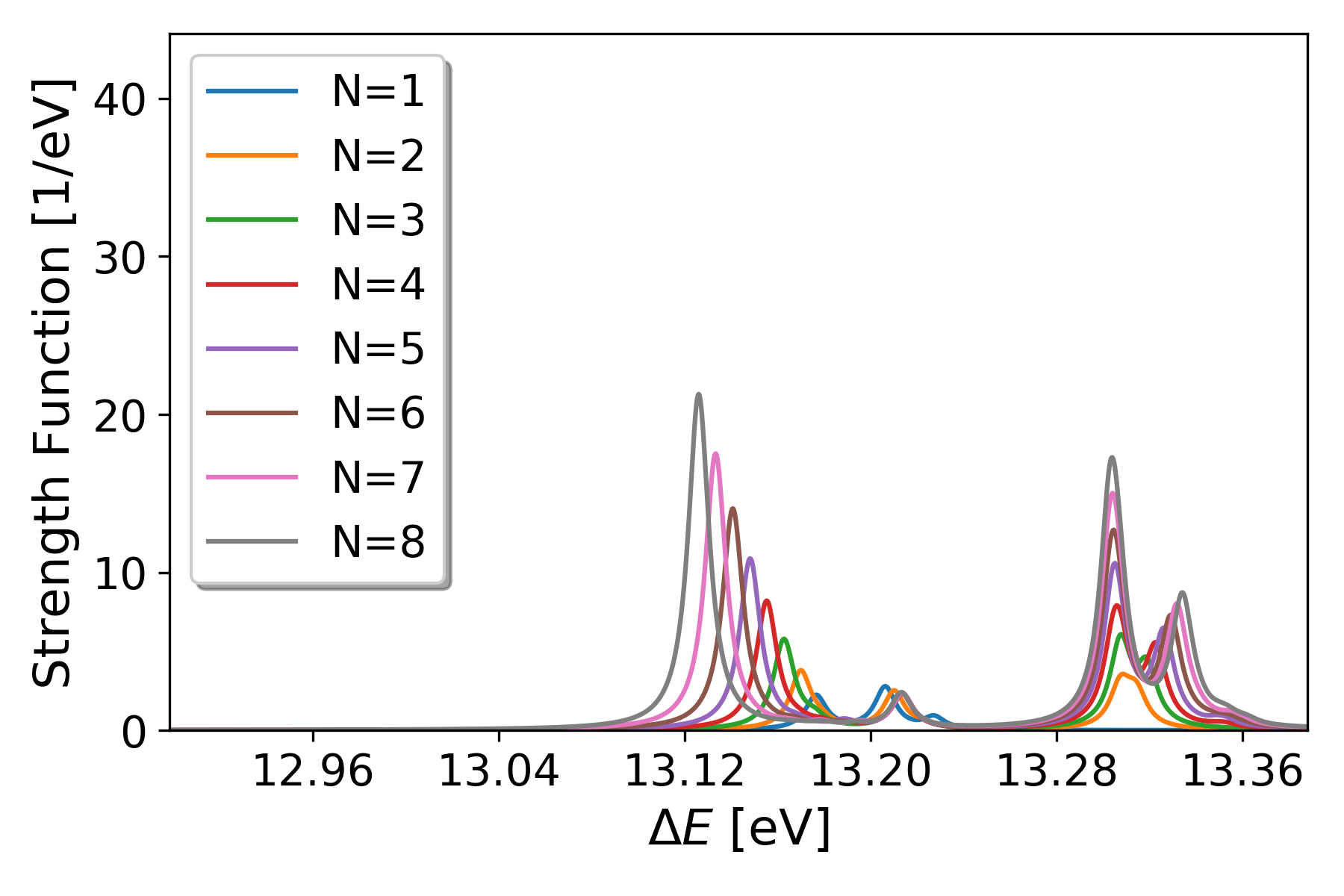}
    \label{fig:}
\end{subfigure}
\caption{Nitrogen dimer chain of variable size $N\in \{1..9\}$ with one perturbed dimer. Cavity tuned to perturbed frequency $\omega_p$ (!) and $\lambda=0.005$. From top to bottom: $\boldsymbol{\lambda}=0$, $\boldsymbol{\lambda}\parallel \bold{e}_x$, $\boldsymbol{\lambda}\parallel \bold{e}_y$ and $\boldsymbol{\lambda}\parallel \bold{e}_z$.
}
\end{figure}

\begin{figure}[H]
\begin{subfigure}{1\textwidth}
\centering
    \includegraphics[width=1\linewidth]{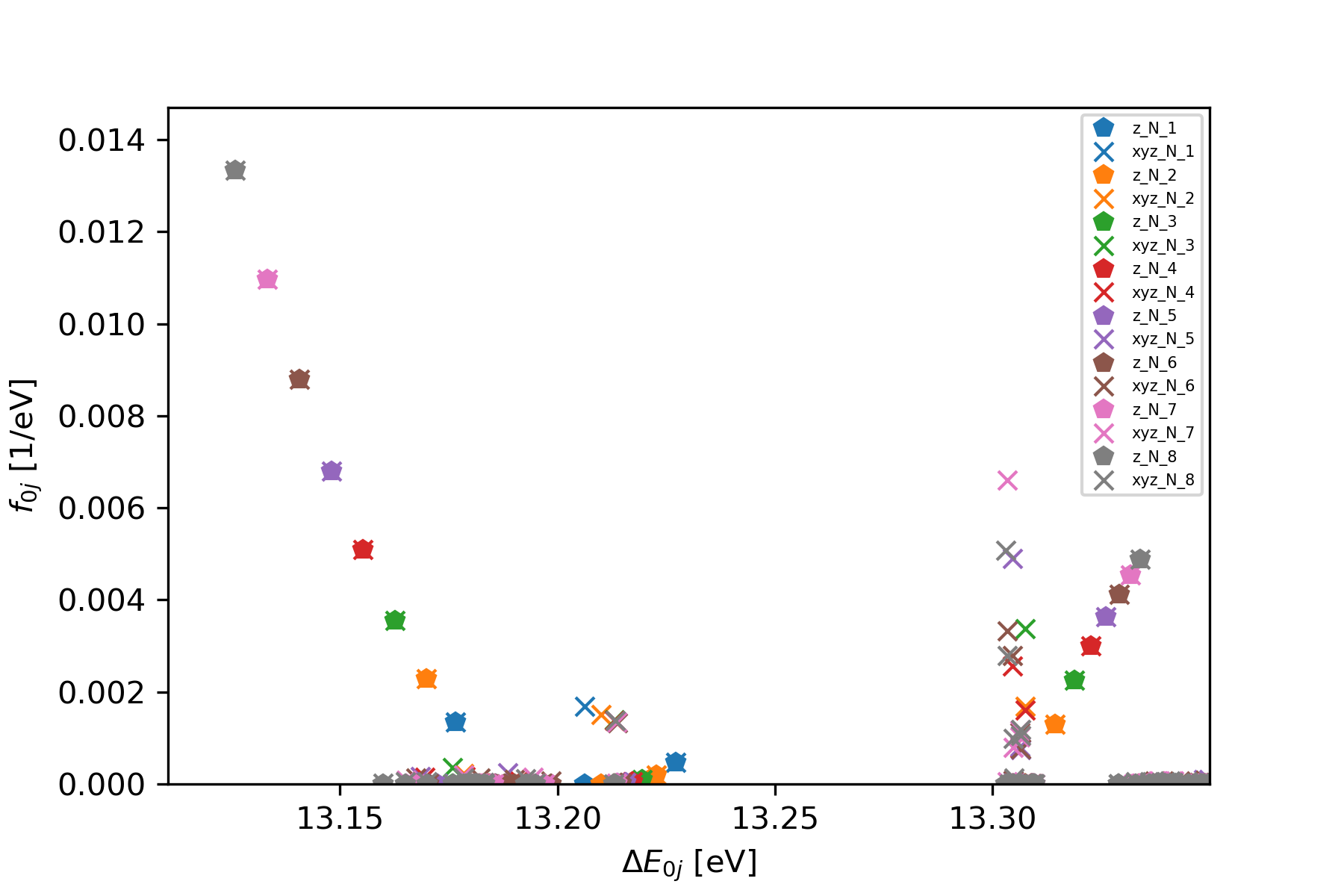}
\end{subfigure}
    \caption{Oscillator strengths for perturbed dimer chain of variable size, with cavity oriented along $z$-axis in coupling regime \RomanNumeralCaps{2}. Cavity tuned to perturbed frequency $\omega_p$ (!) and $\lambda=0.005$. Bold symbols indicate only contributions from transition dipolements along $z$ whereas crosses account equally weighted for all three transition dipole moments along $x,y,z$. The later acts as a basis for the previously shown Lorentz-broadened spectra.}
\label{fig:middle_p_scal}
\end{figure}

\subsubsection{Local Properties}

\begin{figure}[H]
\begin{subfigure}{1\textwidth}
\centering
    \includegraphics[trim=60 0 0 0, clip, width=0.9\linewidth]{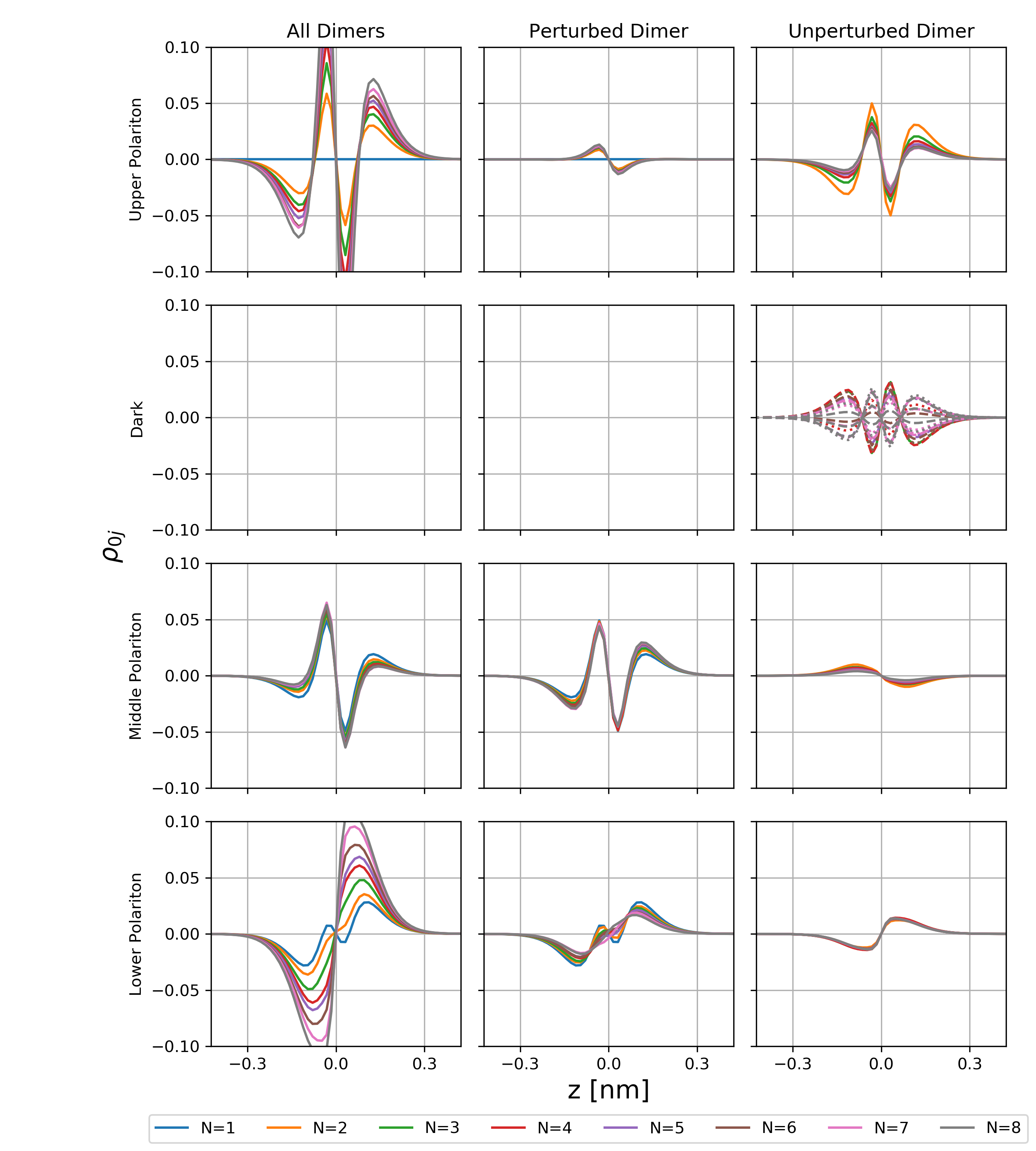}
\end{subfigure}
    \caption{Globally (left column) and locally resolved transition densities projected onto the $z$-axis for different chain lengths $N$ in coupling regime \RomanNumeralCaps{2}. Cavity tuned to perturbed frequency $\omega_p$ (!) and $\lambda=0.005$. For each of the four energy windows (rows), integrated quantities are displayed, except for the dark states. The integration cleans the data and contributes only very little to the overall results. }
\label{fig:middle_p_scal}
\end{figure}

\begin{figure}[H]
\begin{subfigure}{1\textwidth}
\centering
    \includegraphics[trim=60 0 0 0, clip, width=1\linewidth]{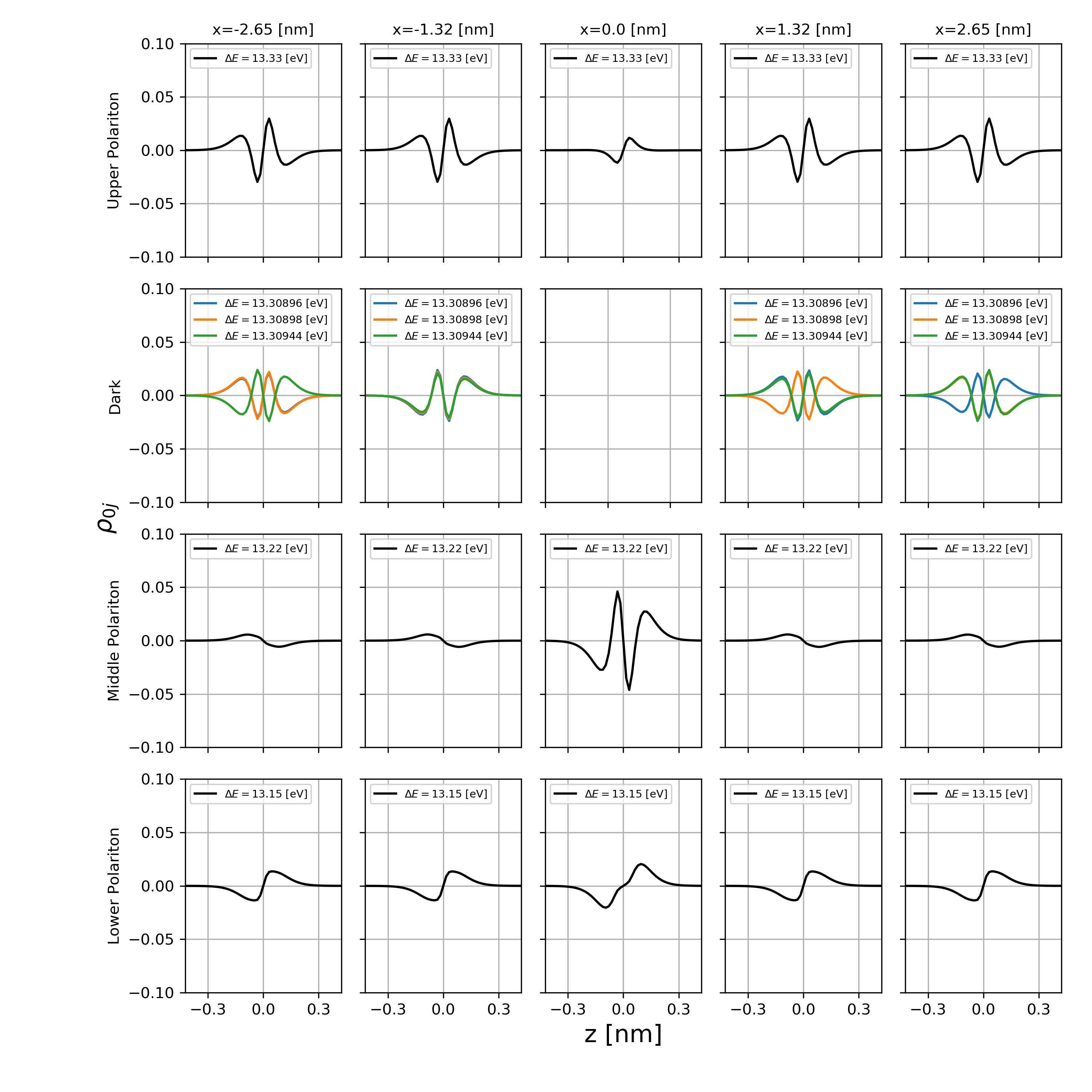}
\end{subfigure}
    \caption{Locally resolved transition densities of each dimer projected onto the $z$-axis for $N=5$ and coupling regime \RomanNumeralCaps{2}. Cavity tuned to perturbed frequency $\omega_p$ (!) and $\lambda=0.005$. For each of the four energy windows (rows), integrated quantities are displayed, except for the three emerging dark states.}
\label{fig:td_loc_Nd_5}
\end{figure}

\begin{figure}
\begin{subfigure}{1\textwidth}
\centering
    \includegraphics[width=0.4\linewidth]{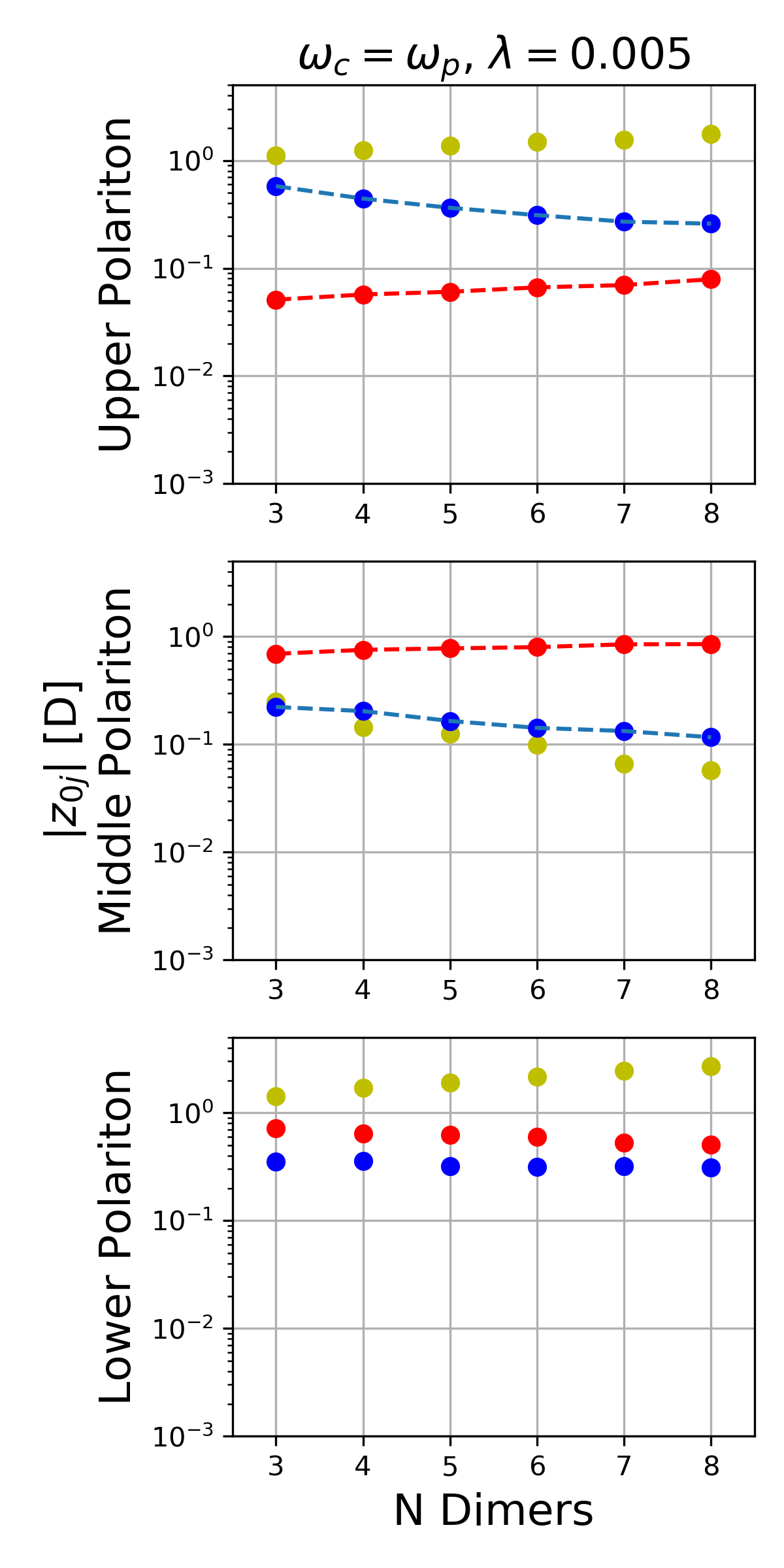}
\end{subfigure}
    \caption{Collective (yellow) and local transition dipole moment scaling with respect to different chain lengths $N$ for the perturbed (red) and unperturbed (blue) dimers in coupling regime \RomanNumeralCaps{2}. The cavity is tuned on $\omega_p$. Special cases $N=\{1,2\}$ are excluded for the sake of clarity (either does not include an unperturbed dimer or no dark states can form). Opposing local scaling behaviour is indicated by blue and red lines. In contrast to the cavity tuned on the unperturbed dimers $\omega_p$, we observe here opposite scaling behaviour also in the upper polaritonic branch and not only for the middle polaritonic branch. }
\label{fig:dipole_scale}
\end{figure}
\section{Simulation Results for Coupling Regime \RomanNumeralCaps{3} ($\lambda=0.01$)}

\subsection{Absorption Spectra for Cavity in Resonance with Unperturbed Dimers}

\begin{figure}[H]
\begin{subfigure}{.8\textwidth}
\centering
    \includegraphics[width=0.9\linewidth]{figures_SI/Spec_pert_D_para_y_ddist_25b_casida_N.png}
    \label{fig:}
\end{subfigure}
\begin{subfigure}{.8\textwidth}
\centering
    \includegraphics[width=0.9\linewidth]{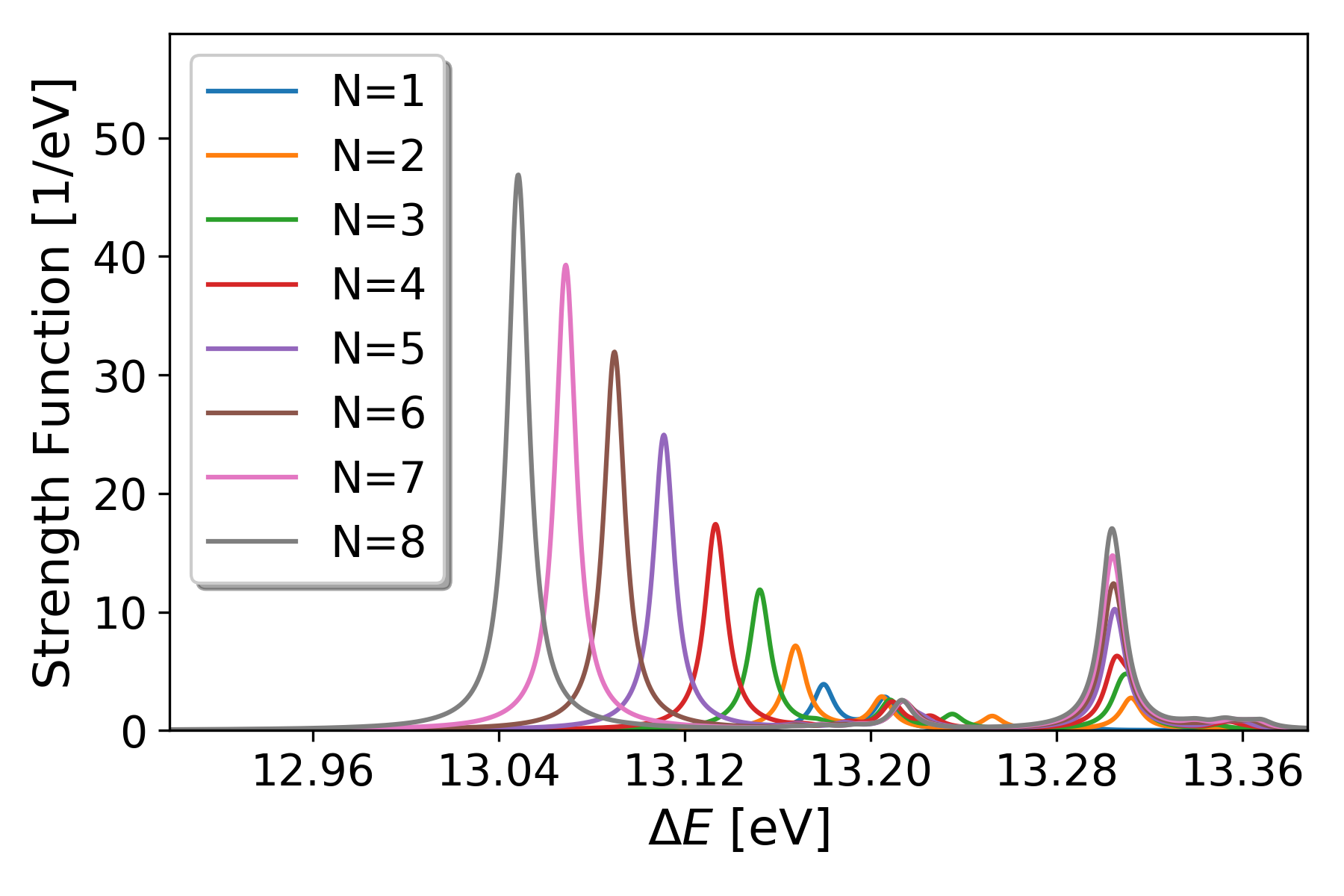}
    \label{fig:}
\end{subfigure}
\caption{Nitrogen dimer chain of variable size $N\in \{1..9\}$ with one perturbed dimer. From top to bottom: $\boldsymbol{\lambda}=0$ and $\boldsymbol{\lambda}\parallel \bold{e}_z$.
}
\end{figure}

\begin{figure}[H]
\begin{subfigure}{1\textwidth}
\centering
    \includegraphics[width=1\linewidth]{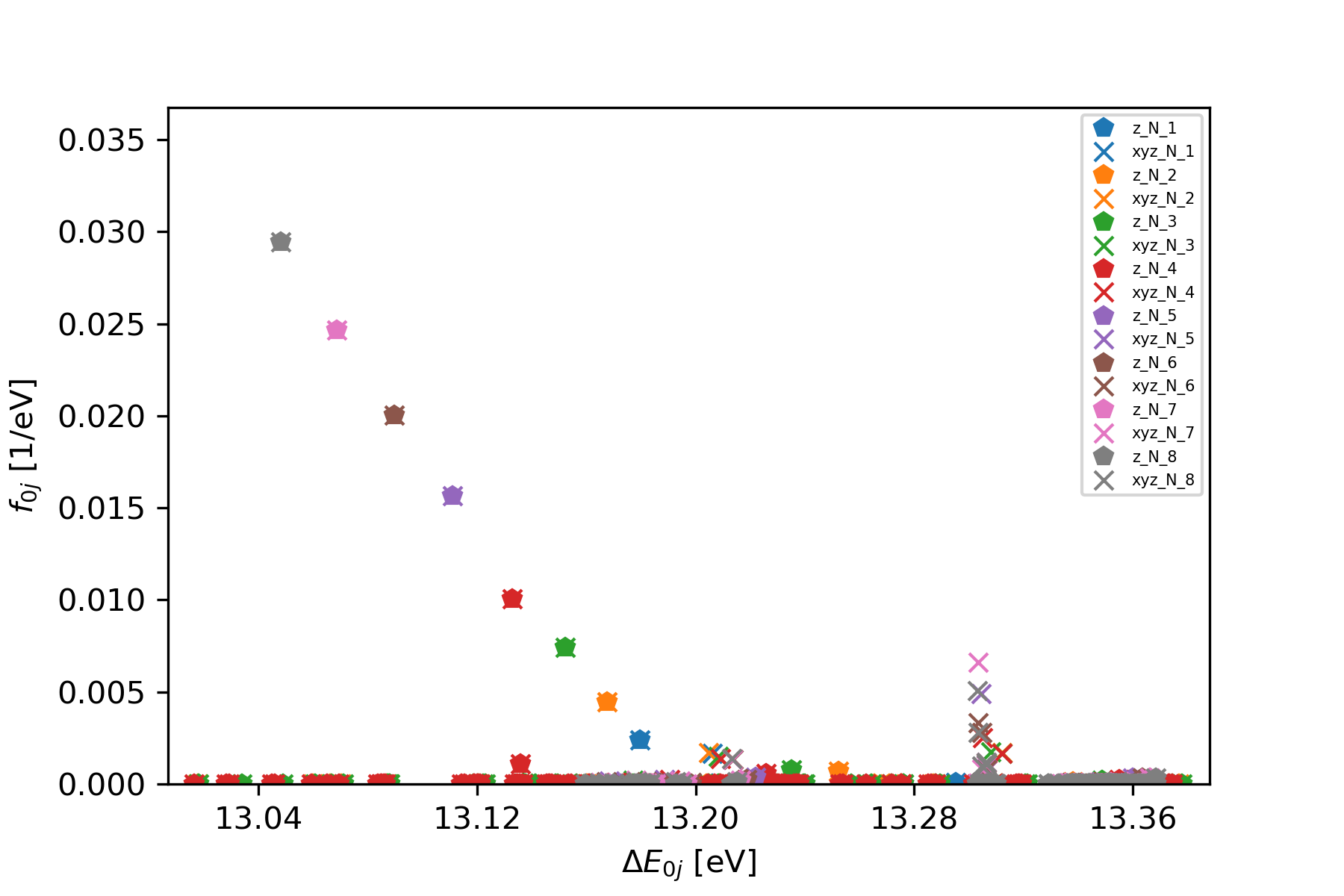}
\end{subfigure}
    \caption{Oscillator strengths for perturbed dimer chain of variable size, with cavity oriented along $z$-axis in coupling regime \RomanNumeralCaps{3}. Bold symbols indicate only contributions from transition dipolements along $z$ whereas crosses account equally weighted for all three transition dipole moments along $x,y,z$. The later acts as a basis for the previously shown Lorentz-broadened spectra.}
\label{fig:middle_p_scal}
\end{figure}

\subsubsection{Local Properties}

\begin{figure}[H]
\begin{subfigure}{1\textwidth}
\centering
    \includegraphics[trim=60 0 0 0, clip, width=0.9\linewidth]{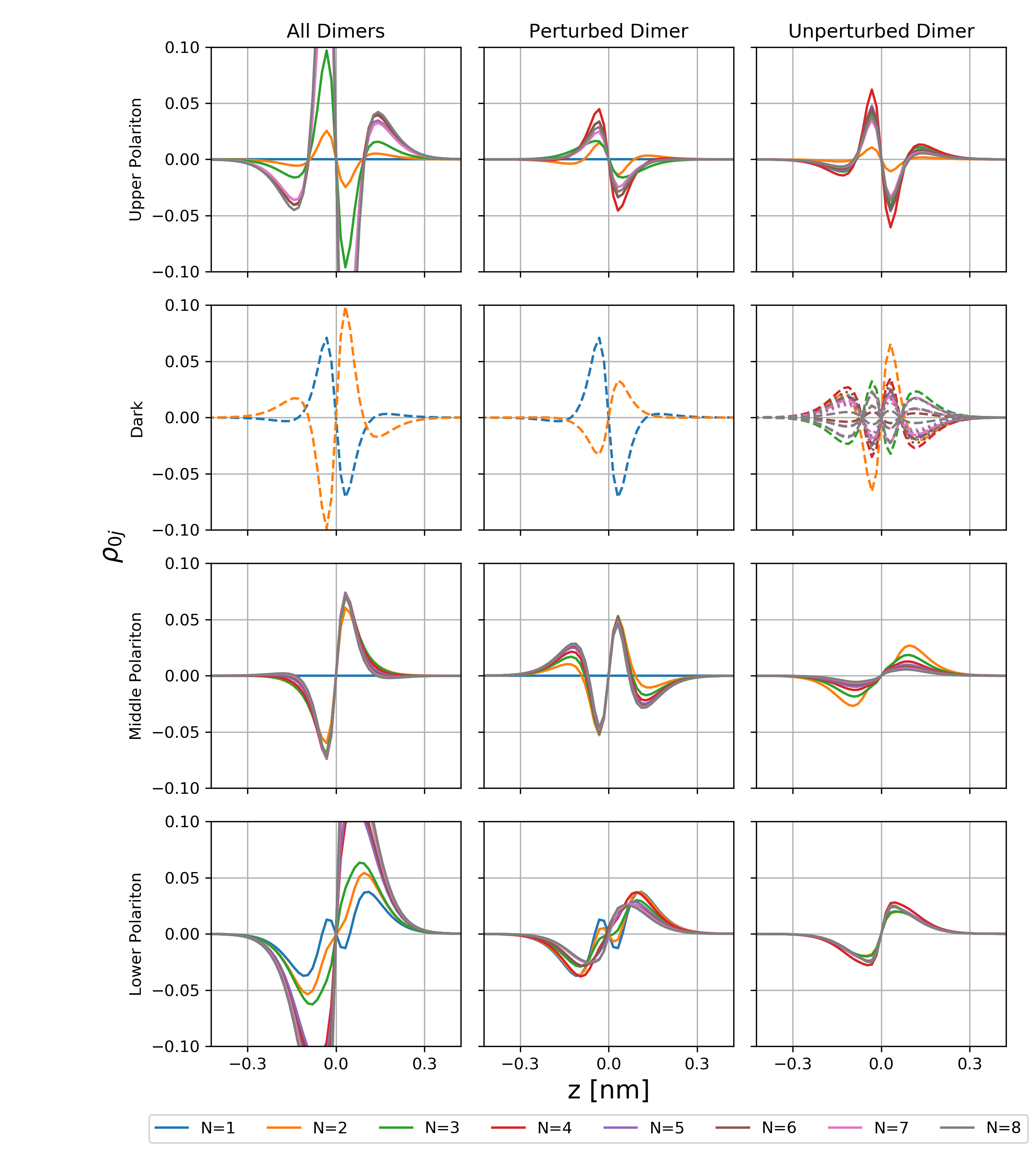}
\end{subfigure}
    \caption{Globally (left column) and locally resolved transition densities projected onto the $z$-axis for different chain lengths $N$ in coupling regime \RomanNumeralCaps{3}. For each of the four energy windows (rows), integrated quantities are displayed, except for the dark states. The integration cleans the data and contributes only very little to the overall results. }
\label{fig:middle_p_scal}
\end{figure}

\begin{figure}[H]
\begin{subfigure}{1\textwidth}
\centering
    \includegraphics[trim=60 0 0 0, clip, width=1\linewidth]{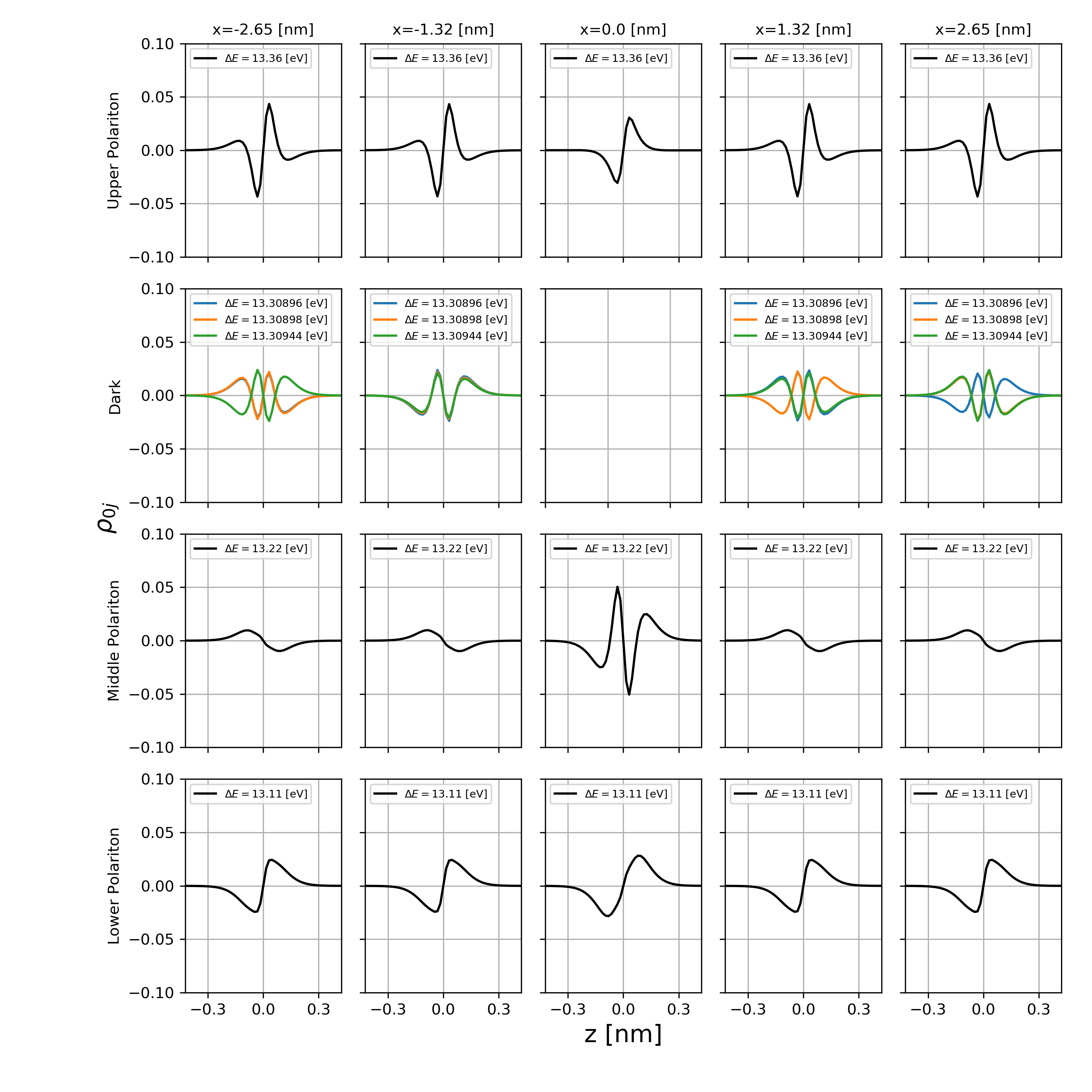}
\end{subfigure}
    \caption{Locally resolved transition densities of each dimer projected onto the $z$-axis for $N=5$ and coupling regime \RomanNumeralCaps{3}. For each of the four energy windows (rows), integrated quantities are displayed, except for the three emerging dark states.}
\label{fig:td_loc_Nd_5}
\end{figure}

